\documentclass[cernpreprint, atlasdraft=false, UKenglish, texmf, orcidlogo]{atlasdoc}
 
\usepackage{atlaspackage}
\usepackage{atlasbiblatex}
 
\usepackage{atlasphysics}

\graphicspath{{logos/}{figures/Public/}}
 
\usepackage{ANA-HDBS-2021-10-PAPER-defs}
 
\usepackage{longtable}

\AtlasTitle{Studies of new Higgs boson interactions through nonresonant $HH$ production in the $b\bar{b}\gamma\gamma$ final state in $pp$ collisions at
$\sqrt{s}=13$~TeV with the ATLAS detector}
 
\AtlasAbstract{
A search for nonresonant Higgs boson pair production in the
$b\bar{b}\gamma\gamma$ final state is performed using 140 fb$^{-1}$ of
proton--proton collisions at a centre-of-mass energy of 13~TeV recorded
by the ATLAS detector at the CERN Large Hadron Collider.
This analysis supersedes and expands upon the
previous nonresonant ATLAS results in this final state based on the
same data sample.
The analysis strategy is optimised to probe anomalous values not only of the
Higgs ($H$) boson self-coupling modifier $\kappa_\lambda$ but also
of the quartic $HHVV$ ($V=W,Z$) coupling modifier $\kappa_{2V}$.
No significant excess above the expected background from
Standard Model processes is observed. An observed upper limit
$\mu_{HH}<4.0$ is set at 95\% confidence level on the Higgs boson
pair production cross-section normalised to its Standard Model prediction.
The 95\% confidence intervals for the coupling modifiers
are $-1.4<\kappa_\lambda<6.9$ and $-0.5<\kappa_{2V}<2.7$,
assuming all other Higgs boson couplings
except the one under study are fixed to the Standard Model predictions.
The results are interpreted in the Standard Model effective field
theory and Higgs effective field theory frameworks in terms of
constraints on the couplings of anomalous Higgs boson
(self-)interactions.
}
 
\author{The ATLAS Collaboration}
 
\AtlasRefCode{HDBS-2021-10}
 
\PreprintIdNumber{CERN-EP-2023-206}

\arXivId{2310.12301}

\AtlasJournal{JHEP}
\AtlasJournalRef{JHEP 01, 066 (2024)}
\AtlasDOI{10.1007/JHEP01(2024)066}

\hypersetup{pdftitle={ATLAS document},pdfauthor={The ATLAS Collaboration}}
 
\begin{document}
 
\maketitle
 
\tableofcontents

\section{Introduction}
 
Since the discovery of a Higgs boson ($H$) in 2012~\cite{HIGG-2012-27,
CMS-HIG-12-028}, the ATLAS and CMS
collaborations at the Large Hadron Collider (LHC) have pursued
an intense programme of measurements of its properties.
All results obtained so far, such as the
spin~\cite{HIGG-2013-17, CMS-HIG-14-018},
intrinsic width~\cite{ATLAS:2023dnm, CMS-HIG-21-013},
production and decay rates~\cite{HIGG-2021-23,CMS-HIG-22-001}
of this particle, are consistent with the predictions of the
Standard Model (SM)~\cite{Englert:1964et,
Higgs:1964ia,Higgs:1964pj,Guralnik:1964eu,Higgs:1966ev,Kibble:1967sv,Glashow:1961tr,Salam:1968rm,Weinberg:1967tq}
for a Higgs boson with an observed mass $m_H$ near
$125~\GeV$~\cite{HIGG-2014-14,ATLAS-CONF-2023-037}.
 
The current measurements provide constraints on the strengths of
the couplings of the Higgs boson to the heaviest of the SM elementary
particles, and on the Higgs boson mass $m_H$.
The latter is one of the parameters of the Higgs boson potential
$V(H) = \frac{1}{2}m_H^2 H^2 + \lambda_{HHH}vH^3 + \frac{1}{4}\lambda_{HHHH}H^4$,
where $v\approx 246$~\GeV\ is the vacuum expectation value of the Higgs field.
Among the SM predictions of the Higgs sector that still remain to be
verified are those for the coupling strengths of the interactions
involving multiple Higgs bosons, such as the trilinear and quartic
Higgs boson self-couplings, $\lambda_{HHH}$ and $\lambda_{HHHH}$, as
well as the quartic couplings between two Higgs bosons and two $W$ or $Z$
bosons, $g_{HHVV}$ $(V=W,Z)$.
In the SM, the trilinear and quartic self-couplings have the value
$\lambda_{HHH}^\mathrm{SM} = \lambda_{HHHH}^\mathrm{SM} = m_H^2/2v^2$, while
the couplings $g_{HHVV}$ are related to the $HWW$ and $HZZ$
couplings $g_{HVV}$ through $g_{HHVV}^\mathrm{SM} = g_{HVV}^\mathrm{SM}/2v$.
 
\begin{figure}[tbh!]
\centering
\subfloat[Trilinear coupling]{
\label{fig:feyn_ggf_tri}
\includegraphics[width=0.43\textwidth]{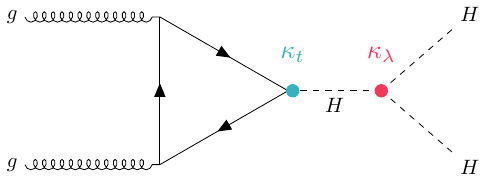} }
\quad\quad
\subfloat[Box diagram]{
\label{fig:feyn_ggf_box}
\includegraphics[width=0.43\textwidth]{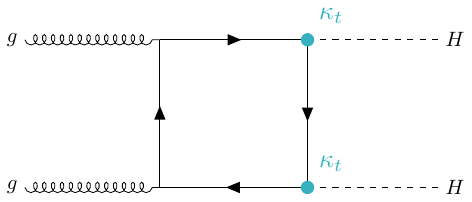} }
\caption[Feynman diagrams of gluon--gluon fusion di-Higgs boson production.]{
The Feynman diagrams for the dominant gluon--gluon fusion
production processes. In the SM, the (a) trilinear
coupling process, (b) box diagram, and the destructive
interference between the two processes, contribute to the total
cross-section. In the figure, \klambda{} represents the Higgs
boson trilinear coupling modifier. The quark content in the
diagram is dominated by the top-quark contribution due to the
large top-quark Yukawa coupling to the Higgs boson. The
corresponding coupling strength modifier is denoted by $\kappa_t$.
}
\end{figure}

\begin{figure}[tbh!]
\centering
\subfloat[Trilinear coupling]{
\label{fig:feyn_vbf_hhh}
\includegraphics[height=0.2\textwidth]{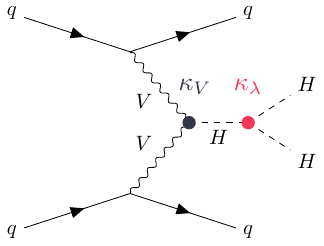} }
\quad
\subfloat[$HHVV$ vertex]{
\label{fig:feyn_vbf_vvhh}
\includegraphics[height=0.2\textwidth]{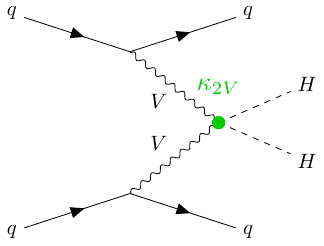} }
\quad
\subfloat[$VVH$ Production mode]{
\label{fig:feyn_vbf_vvh}
\includegraphics[height=0.2\textwidth]{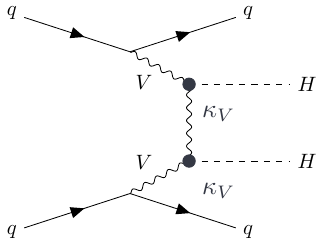} }
\caption[Feynman diagrams of Vector Boson Fusion di-Higgs production.]{
The VBF production of Higgs boson pairs via
(a) the trilinear coupling,
(b) the $HHVV$ vertex,
and (c) the $VVH$ production
mode.
In the figure, $\kappa_V$ and $\kappa_{2V}$ denote the
$HVV$ and $HHVV$ coupling strength modifiers.
}
\end{figure}

A significant effort has been dedicated by the ATLAS and CMS
collaborations to search for processes that are particularly sensitive to $\lambda_{HHH}$ and $g_{HHVV}$, such as Higgs boson pair production
in gluon--gluon fusion (ggF) and vector-boson fusion (VBF).
In the SM, ggF $HH$ production proceeds through the destructive interference
of two leading Feynman diagrams: one for
the process $gg\to H^* \to HH$, involving an intermediate virtual Higgs
boson $(H^*)$ and a $HHH$ vertex (Figure~\ref{fig:feyn_ggf_tri}),
and a second one describing a loop-mediated process in which two Higgs
bosons are radiated off a virtual quark (Figure~\ref{fig:feyn_ggf_box}).
VBF $HH$ production is induced at tree level in the SM by three
Feynman diagrams in which the two vector bosons radiated by the
scattering quarks either fuse into a virtual Higgs boson $H^*$
decaying into two Higgs bosons via a $HHH$ vertex
(Figure~\ref{fig:feyn_vbf_hhh}), fuse into two Higgs bosons via a
$HHVV$ vertex (Figure~\ref{fig:feyn_vbf_vvhh}), or produce two Higgs bosons
via $t$-channel scattering through two $HVV$ interactions
(Figure~\ref{fig:feyn_vbf_vvh}).
The amplitudes of diagrams involving a $HHH$ vertex are
proportional to $\lambda_{HHH}$, while those of diagrams involving a $HHVV$
vertex are proportional to $g_{HHVV}$.
For this reason, the results of the searches for $HH$ production can be used
to infer the values of the coupling modifiers
$\kappa_\lambda \equiv \lambda_{HHH}/\lambda_{HHH}^\mathrm{SM}$
and $\kappa_{2V} = g_{HHVV}/g_{HHVV}^\mathrm{SM}$.
An observed value of these coupling modifiers
significantly different from unity would provide a proof of
non-SM Higgs boson interactions~\cite{Agrawal_2020}.
 
In the SM, these processes are expected to be rare,
with cross-sections that are about three orders of magnitude smaller
than those of single Higgs boson production:
$\sigma^{HH}_\ggF = 31.1^{+2.1}_{-7.2}$~fb~\cite{Grazzini:2018bsd,
Dawson:1998py, Borowka:2016ehy,
Baglio:2018lrj, deFlorian:2013jea, Shao:2013bz,
deFlorian:2015moa, Baglio:2020wgt}
and $\sigma^{HH}_\VBF=1.73\pm 0.04$~fb~\cite{Dreyer:2018,Baglio:2013,Ling:2014}
for $m_H=125$~\GeV and a proton-proton
centre-of-mass energy of $\sqrt{s}=13$~\TeV.
It is thus crucial to analyse the latest available data sample and
reconstruct as many decay final states of the Higgs boson pairs
as possible.
The most stringent constraints on \kl and \kvv
exploit the entire sample of proton--proton ($pp$) collisions
provided by the LHC during its second phase of data-taking (Run~2, 2015--2018),
and a multitude of Higgs boson decay channels.
In particular, the ATLAS experiment recently released the results of searches
based on the full Run~2 data in the three most sensitive channels,
\bbyy~\cite{HDBS-2018-34_custom}, \bbtt~\cite{HDBS-2018-40},
and \bbbb~\cite{ATLAS4b}, and their combination~\cite{HDBS-2022-03_custom}.
No excess over the SM background was observed, and constraints on the coupling modifiers \kl and \kvv were set at the 95\% confidence
level (CL). The observed (expected for $\kl=1$) 95\% confidence interval for
\kl\ when all other coupling strength modifiers are set to unity is
$-0.6 < \kl < 6.6$ ($-2.1 < \kl < 7.8$) after combining the three
$HH$ decay channels.
For \kvv, the observed (expected) 95\% confidence interval when all
other coupling strength modifiers are set to unity is
$0.1< \kvv < 2.0$ ($0.0 < \kvv < 2.1$).
With a similar data sample, CMS also reported similar results in their
\bbyy~\cite{CMS-HIG-19-018}, \bbtt~\cite{CMS-HIG-20-010}, and
\bbbb~\cite{CMS-HIG-20-005,CMS-B2G-22-003} channels,
observing 95\% CL intervals of $-1.2<\kl<6.5$ and $0.7<\kvv<1.4$,
based on a different statistical procedure~\cite{CMS-HIG-22-001}.
 
The \bbyy final state has an expected branching ratio (0.26\%) that is
significantly smaller than that of \bbbb (34\%) and \bbtt (7.3\%).
However, the larger expected signal-to-background (S/B) ratio and the
higher trigger efficiency and thus larger acceptance in phase-space
regions (e.g., at small values of
the $HH$ invariant mass), where potential deviations from the SM are
expected to be enhanced, compensate for the lower expected event yield
and lead to a sensitivity similar to that
of the other two decay modes.
The latest results for \hhyybb with the full Run~2 ATLAS data,
based on the event selection and classification of Ref.~\cite{HDBS-2018-34_custom}
but using the statistical procedures of Ref.~\cite{HDBS-2022-03_custom} and of this analysis,
yields the following observed (expected) one-dimensional 95\% confidence intervals:
$-1.4<\kl<6.5$ ($-3.2<\kl<8.1$) and $-0.8<\kvv<3.0$ ($-1.6<\kvv<3.7$).
 
This paper presents an updated search for nonresonant Higgs boson pair
production in the \bbyy final state using the full Run~2 ATLAS data,
superseding and expanding upon the nonresonant results of
Ref.~\cite{HDBS-2018-34_custom}.
Compared to the previous publication, an identical
event selection and a similar
analysis strategy are used, but a reoptimised classification of events
in categories with different S/B leads to a higher sensitivity to
the \kl and \kvv coupling modifiers.
The new event classification relies on improved multivariate classifiers,
also exploiting the kinematic features of VBF $HH$ production for SM
and anomalous values of $\kl$ and $\kvv$.
After the events are classified in mutually orthogonal event categories,
the signal cross-section is estimated through a simultaneous
maximum-likelihood fit to the diphoton invariant mass spectrum of the
selected events in each category.
The fit probes an enhancement in event yields around the experimental
value of the Higgs boson mass over the predicted background,
consisting of the sum
of a monotonically decreasing distribution from continuum photon and jet
production and a peak from singly produced Higgs bosons
decaying into two photons.
 
Another novelty compared to the previous publication is the
interpretation of the results in two effective field theory (EFT)
extensions to the SM, the Higgs effective field theory
(HEFT)~\cite{Alonso:2012px,Buchalla:2013rka} and
the SM effective field theory
(SMEFT)~\cite{Buchmuller:1985jz,Brivio:2017vri}.
The data are used to set constraints on the Wilson coefficients of
operators of the EFT Lagrangians describing anomalous Higgs boson
interactions in both frameworks.
HEFT and SMEFT describe the same effective interactions, but
with different bases of operators. One advantage of HEFT compared with SMEFT
is that it provides a one-to-one relation between operators (and
corresponding Wilson coefficients) and effective interactions, which
allows single- and di-Higgs boson couplings to be separated, leading to
simplified $HH$ interpretations.
 
This paper is organised as follows.
Section~\ref{sec:atlas} describes the experimental apparatus.
The data and simulated event samples used for the measurements
are summarised in Section~\ref{sec:datamc}.
Section~\ref{sec:selection} is devoted to the event selection
and classification, with an emphasis on the novelties of the latter
compared to the previous publication.
The \myy signal and background models used in the final fit are described in
Section~\ref{sec:sb_modelling}.
The systematic uncertainties in the measurement and the results
are given in Sections~\ref{sec:systematics} and~\ref{sec:results},
respectively.
Finally, the procedure and results of the EFT interpretation are detailed
in Section~\ref{sec:EFT}.
Section~\ref{sec:conclusion} provides the conclusions.


\section{The ATLAS detector}
\label{sec:atlas}
 
\newcommand{\AtlasCoordFootnote}{
ATLAS uses a right-handed coordinate system with its origin at the nominal interaction point (IP)
in the centre of the detector and the \(z\)-axis along the beam pipe.
The \(x\)-axis points from the IP to the centre of the LHC ring,
and the \(y\)-axis points upwards.
Cylindrical coordinates \((r,\phi)\) are used in the transverse plane,
\(\phi\) being the azimuthal angle around the \(z\)-axis.
The pseudorapidity is defined in terms of the polar angle \(\theta\) as \(\eta = -\ln \tan(\theta/2)\).
The rapidity $y$ is defined in terms of the energy, the momentum and the polar angle \(\theta\):
$y=\frac{1}{2} \ln  \left( \frac{E+p \cdot \cos \theta}{E-p \cdot \cos \theta} \right)$.
The angular distance is measured in units of \(\Delta R \equiv \sqrt{(\Delta\eta)^{2} + (\Delta\phi)^{2}}\).}
 
The ATLAS detector~\cite{PERF-2007-01} at the LHC covers nearly the entire solid angle around the collision point.\footnote{\AtlasCoordFootnote}
It consists of an inner tracking detector surrounded by a thin superconducting solenoid, electromagnetic and hadron calorimeters,
and a muon spectrometer incorporating three large superconducting air-core toroidal magnets.
 
The inner-detector system (ID) is immersed in a \qty{2}{\tesla} axial magnetic field
and provides charged-particle tracking in the range of \(|\eta| < 2.5\).
The high-granularity silicon pixel detector covers the vertex region and typically provides four measurements per track,
the first hit normally being in the insertable B-layer (IBL) installed before Run~2~\cite{ATLAS-TDR-19,PIX-2018-001}.
It is followed by the silicon microstrip tracker (SCT), which usually provides eight measurements per track.
These silicon detectors are complemented by the transition radiation tracker (TRT),
which enables radially extended track reconstruction up to \(|\eta| = 2.0\).
The TRT also provides electron identification information
based on the fraction of hits (typically 30 in total) above a higher energy-deposit threshold corresponding to transition radiation.
 
The calorimeter system covers the pseudorapidity range of \(|\eta| < 4.9\).
In the region \(|\eta|< 3.2\), electromagnetic calorimetry is provided by barrel and
endcap high-granularity lead/liquid-argon (LAr) calorimeters,
with an additional thin LAr presampler covering \(|\eta| < 1.8\)
to correct for energy loss in material upstream of the calorimeters.
Hadron calorimetry is provided by the steel/scintillator-tile calorimeter,
segmented into three barrel structures with \(|\eta| < 1.7\), and two copper/LAr hadron endcap calorimeters.
The solid angle coverage is completed with forward copper/LAr and tungsten/LAr calorimeter modules
optimised for electromagnetic and hadronic energy measurements respectively.
 
The muon spectrometer (MS) comprises separate trigger and
high-precision tracking chambers measuring the deflection of muons in a magnetic field generated by the superconducting air-core toroidal magnets.
The field integral of the toroids ranges between \num{2.0} and \qty{6.0}{\tesla\metre}
across most of the detector.
Three layers of precision chambers, each consisting of layers of monitored drift tubes, cover the region \(|\eta| < 2.7\),
complemented by cathode-strip chambers in the forward region, where the background is highest.
The muon trigger system covers the range of \(|\eta| < 2.4\) with resistive-plate chambers in the barrel, and thin-gap chambers in the endcap regions.
 
Interesting events are selected by the first-level trigger system implemented in custom hardware,
followed by selections made by algorithms implemented in software in the high-level trigger~\cite{TRIG-2016-01}.
The first-level trigger accepts events from the \qty{40}{\MHz} bunch crossings at a rate below \qty{100}{\kHz},
which the high-level trigger further reduces in order to record events to disk at about \qty{1}{\kHz}.
 
An extensive software suite~\cite{ATL-SOFT-PUB-2021-001} is used in data simulation, in the reconstruction
and analysis of real and simulated data, in detector operations, and in the trigger and data acquisition
systems of the experiment.


\section{Data and simulation samples}
\label{sec:datamc}
 
The measurements presented in this paper use $pp$ collision
data collected by the ATLAS experiment during the LHC
Run~2 at $\sqrt s =13$~\TeV.
After 
data quality requirements~\cite{DAPR-2018-01}, the
integrated luminosity of the data sample is
$140.1 \pm 1.2$~fb$^{-1}$~\cite{DAPR-2021-01}.
 
The simulated event samples used in this study are summarised in
Table~\ref{tab:mcsamples}. Besides the samples already used in
Ref.~\cite{HDBS-2018-34_custom}, VBF $HH$ samples were produced
for additional $\kappa_{2V}$ and $\kappa_V$ variations
(where $\kv = g_{HVV}/g_{HVV}^\mathrm{SM}$ is the
$HVV$ coupling modifier), and a
dedicated diphoton + two $b$-jet sample was generated.
 
\begin{table}[ht!]
\centering
\caption{Summary of the nominal Higgs boson pair signal,
single Higgs boson background
and continuum background event samples used in this analysis.
The generator used in the simulation, the parton distribution
function (PDF) set, and the set of tuned
parameters (tune) are also provided.
The final two columns list the accuracy in QCD of the event
generator and the order in QCD of the calculated cross-section
for the $HH$ signal and the single Higgs boson background
(LO: leading order, NLO: next-to-leading order,
NNLO: next-to-next-to-leading order, N$^3$LO:
next-to-next-to-next-to-leading order).
More details are given in
the text and in Ref.~\cite{HDBS-2018-34_custom}.
The accuracy and cross-sections for the nonresonant background processes
are omitted since their shape parameters and overall normalisation
are determined from fits to the data.
}
\label{tab:mcsamples}
\resizebox{\columnwidth}{!}{
\begin{tabular}{ lllllll }
\toprule
Process & Generator    & PDF set    & Showering & Tune & Accuracy & Order of \\
&              &            &           &      &          & $\sigma$ calculation  \\
\midrule
ggF $HH$  & \POWHEGBOX[v2]~\cite{Nason:2004rx, Frixione:2007vw, Alioli:2010xd,powheg-sm,Heinrich:2019bkc} & \PDFforLHC[15nlo]~\cite{Butterworth:2015oua}  &   \PYTHIA[8.2]~\cite{Sjostrand:2014zea} &  A14~\cite{ATL-PHYS-PUB-2014-021}  & NLO & NNLO     \\
VBF $HH$  & \MGNLO~\cite{Alwall:2014hca} & \NNPDF[3.0nlo]~\cite{Ball:2014uwa} &  \PYTHIA[8.2] &  A14 & LO & N$^{3}$LO \\
\midrule
ggF $H$  & NNLOPS~\cite{Nason:2004rx, Frixione:2007vw, Alioli:2010xd, Hamilton:2013fea, Hamilton:2015nsa}  & \PDFforLHC[15nlo]&   \PYTHIA[8.2]  &  AZNLO~\cite{STDM-2012-23} & NNLO & N$^{3}$LO \\
VBF $H$ & \POWHEGBOX[v2]~\cite{Nason:2004rx,Frixione:2007vw,Alioli:2010xd, Nason:2009ai} & \PDFforLHC[15nlo]  &    \PYTHIA[8.2]     & AZNLO  & NLO & NNLO \\
$WH$ &\POWHEGBOX[v2]~\cite{Nason:2004rx,Frixione:2007vw,Alioli:2010xd, Luisoni:2013kna} & \PDFforLHC[15nlo]  &   \PYTHIA[8.2]      & AZNLO  & NLO & NNLO \\
$qq\to ZH$ & \POWHEGBOX[v2]~\cite{Nason:2004rx,Frixione:2007vw,Alioli:2010xd, Luisoni:2013kna} &  \PDFforLHC[15nlo]  &   \PYTHIA[8.2]      & AZNLO  & NLO & NNLO \\
$gg\to ZH$ &  \POWHEGBOX[v2]~\cite{Nason:2004rx,Frixione:2007vw,Alioli:2010xd, Luisoni:2013kna}  & \PDFforLHC[15nlo]  &   \PYTHIA[8.2]      & AZNLO  & LO & NLO \\
$t\bar{t}H$ & \POWHEGBOX[v2]~\cite{Nason:2004rx,Frixione:2007vw,Alioli:2010xd, Hartanto:2015uka} & \NNPDF[3.0nlo]  &   \PYTHIA[8.2]      &  A14 & NLO & NNLO \\
$b\bar{b}H$ &  \POWHEGBOX[v2]~\cite{Nason:2004rx,Frixione:2007vw,Alioli:2010xd, Jager:2015hka}  & \NNPDF[3.0nlo]  &  \PYTHIA[8.2]   & A14  & NLO & NNLO \\
$tHq$ & \MGNLO &  \NNPDF[3.0nlo]  & \PYTHIA[8.2] &   A14   & NLO & NLO \\
$tHW$  & \MGNLO &  \NNPDF[3.0nlo]  & \PYTHIA[8.2] &  A14   & NLO & NLO \\
\midrule
$\gamma\gamma+$jets &   \SHERPA[2.2.4]~\cite{Bothmann:2019yzt} & \NNPDF[3.0nnlo]  &  \SHERPA[2.2.4] & -- & -- & -- \\
$\gamma\gamma b\bar{b}$ & \SHERPA[2.2.12]~\cite{Bothmann:2019yzt} & \NNPDF[3.0nnlo]  &  \SHERPA[2.2.12] & -- & -- & -- \\
$t\bar{t} \gamma \gamma$ & \MGNLO\  &  \NNPDF[2.3lo] & \PYTHIA[8.2] & A14 & -- & -- \\
\bottomrule
\end{tabular}
}
\end{table}
 
Signal samples consist of simulated events from
nonresonant ggF and VBF production of Higgs boson pairs, with one
Higgs boson decaying into $b\bar{b}$ and the other one into
$\gamma\gamma$.
In addition to the samples in Table~\ref{tab:mcsamples},
a ggF $HH$ sample was generated
with the same settings as the nominal sample but with the non-SM value
of the self-coupling modifier $\kl=10$, and then passed through the
detector simulation and the reconstruction algorithms.
A reweighting technique based on the particle-level invariant
mass $m_{HH}$ of the Higgs boson pair
is applied to the
$\kl=1$ sample to determine the ggF $HH$ signal yield and
kinematic distributions for any value of $\kl$~\cite{ATL-PHYS-PUB-2019-007}.
The particle-level $m_{HH}$ spectrum for any generic value of $\kl$ is
calculated from the $m_{HH}$ distributions of three ggF $HH$ samples
generated at particle level for $\kl=0$, 1, and 20.
To determine the potential `non-closure' in the reweighting
process from residual kinematic effects,
the procedure is validated by comparing the predicted event yields
and kinematic distributions of the simulated sample generated with
$\kl=1$ and reweighted to $\kl=10$
with those of the simulated sample generated under the hypothesis $\kl=10$.
Furthermore, 12 additional VBF $HH$ samples were generated
and simulated with the same set-up and settings as the nominal VBF sample
but using non-SM combinations of the coupling strength scale
factors \kl, \kvv and \kv.
A linear combination of a `basis' formed by the SM sample and five
of the other 12 samples, corresponding to the combinations of
the \kl, \kvv, and \kv couplings
(1, 1.5, 1), (0, 1, 1), (10, 1, 1), (1, 3, 1), ($-5$, 1, 0.5),
is used to determine the expected yields
and distributions for any value of \kl, \kvv, and \kv. The remaining
seven samples are compared with the corresponding
predictions from the interpolation procedure for
validation purposes. The same procedure was used in the
measurements presented in Refs.~\cite{ATLAS4b, HDBS-2022-03_custom}.
 
Background samples include simulated events of single Higgs
bosons decaying into $\gamma\gamma$ produced by ggF, VBF, in association
with a $W$ or $Z$ boson, with a $t\bar{t}$ or $b\bar{b}$ pair, or with
a single top-quark $t$. Simulated event samples of
continuum diphoton production in association with top quark pairs
($t\bar{t}\gamma\gamma$) or with jets from quarks of other flavours
($\gamma\gamma$+jets) were also produced, to optimise the event
classification described in Section~\ref{ssec:categories}.
In addition to the previous samples, shared with
Ref.~\cite{HDBS-2018-34_custom},
a sample of simulated continuum diphoton plus two $b$-jets
events ($\gamma\gamma b\bar{b}$) was generated with
\Sherpa[2.2.12]~\cite{Bothmann:2019yzt}
using NLO matrix elements for the production of the two photons and the
two $b$-quarks in the four-flavour scheme, with additional jets
produced in the parton shower.
Due to the increased efficiency from generator-level requirements on
the $b$-quarks, the use of this new sample reduces the statistical
uncertainty in the main component of the nonresonant background
originating from $\yybb$ events by a factor of two, despite
containing about 60 times fewer simulated events than the inclusive
diphoton sample. This sample was used to study the background diphoton
invariant mass distribution, as described in Section~\ref{sec:sb_modelling}.
 
All generated samples were passed through a detailed simulation
of the ATLAS detector response~\cite{SOFT-2010-01} based on
\GEANT~\cite{Agostinelli:2002hh},
except for the inclusive diphoton sample, which
was interfaced to a fast detector simulation based on a parametric
description of the calorimeter response~\cite{ATL-PHYS-PUB-2010-013},
and for the ggF $HH$ particle-level samples used for the
$m_{HH}$-based reweighting procedure, for which the detector response
was not simulated.
The generation of the simulated event samples includes the effect of
multiple inelastic \(pp\) interactions per bunch crossing,
and the effect on the detector response
of interactions from bunch crossings before or after the one
containing the hard interaction.
The inelastic $pp$ events were generated with \PYTHIA[8.186] using the
\NNPDF[2.3lo] PDF set and the A3 tune~\cite{ATL-PHYS-PUB-2016-017}.
The Higgs boson mass was assumed to be 125~\GeV\ in both simulation and
the analysis of the data. The impacts of the differences
relative to the best-fit values of the $m_H$ measurements reported
in Refs.~\cite{HIGG-2014-14,ATLAS-CONF-2023-037}, and the
effects of the corresponding experimental uncertainties in $m_H$, are
negligible.


\section{Event selection and classification}
\label{sec:selection}
 
The same preselection as described in Ref.~\cite{HDBS-2018-34_custom} is used to suppress
the background while providing good signal efficiency.
It is briefly summarised in Section~\ref{ssec:preselection}.
The selected events are then classified into orthogonal categories based
on multivariate discriminants using several input kinematic quantities.
The definition of the event categories, described in
Section~\ref{ssec:categories}, is chosen in order to optimise
the expected constraints on the coupling modifiers \kl and \kvv.
 
\subsection{Event preselection}
\label{ssec:preselection}

To identify $H \to \gamma\gamma$ decays, events were collected with
diphoton triggers~\cite{TRIG-2018-05} with nominal transverse
momentum (\pT) thresholds of 35~\GeV\ and 25~\GeV\ for the
leading- and subleading-\pT candidates, respectively.
Selected events are required to contain two photon candidates in
the acceptance of the finely segmented part of the electromagnetic
calorimeter ($|\eta|<1.37$ or $1.52<|\eta|<2.37$).
The candidates must be identified as photons by an algorithm
based on the shower shapes reconstructed in the calorimeter.
Of all potential reconstructed collision vertices, the primary diphoton
vertex (PV) is selected by a neural-network algorithm using extrapolated photon
trajectories and tracks associated with the candidate
vertices~\cite{HIGG-2013-08}.
The photon candidates must also meet the requirements
of an isolation algorithm based on the energy flow
in the calorimeter and the total transverse momentum of charged particle tracks
from the PV in the inner detector, in cones surrounding the photon
direction~\cite{EGAM-2018-01}.
The two leading photons passing these selections
are then required to have an invariant mass \myy between
105 and 160~\GeV\ and transverse momenta above 35\% and 25\% of \myy.
 
Jets are reconstructed from
particle-flow objects built from noise-suppressed positive-energy
topological clusters in the calorimeter and reconstructed tracks
using the anti-$k_t$ clustering algorithm with the parameter $R=0.4$~\cite{Cacciari:2011ma,Cacciari:2008gp}.
Jet candidates are required to have $\pT>25$~\GeV\ and $|y|<4.4$.
Jets in the fiducial acceptance of the inner detector ($|\eta|<2.4$)
and with $\pT<60$~\GeV\ must be identified by a `jet-vertex tagger'
as originating from the PV~\cite{ATLAS-CONF-2014-018}.
To target $H \to \bb$ decays, events are required to contain
exactly two {\em $b$-tagged} jets, defined as {\em central} jets
(those in the acceptance of the inner detector ($|\eta|<2.5$))
that satisfy the criteria of the `DL1r' $b$-tagging algorithm with a nominal
efficiency of 77\% for $b$-jets and a misidentification rate of 1/170 (1/5) for
light-flavour (charm) jets in $t\bar{t}$ simulated
events~\cite{FTAG-2019-07_custom}. 
A correction factor is applied to the energy of the two $b$-tagged
jets to account for possible contributions from muons originating from
semileptonic $b$-hadron decays and undetected energy from neutrinos
and out-of-cone effects~\cite{HDBS-2018-34_custom}. Jets failing to satisfy the
$b$-tagging requirement are ranked from first to last based on a
discrete $b$-tagging score defined by three bins, corresponding to
central jets with DL1r efficiencies of 77\%--85\% and 85\%--100\%, and
non-central jets. Jets with the same score are ranked by \pT.

Events with six or more central jets, or with one or more isolated
lepton (electron or muon) candidates with $\pT>10$~\GeV\ and passing
the lepton identification criteria are rejected in order to suppress
background from $t\bar{t}H(\gamma\gamma)$ and inclusive $t\bar{t}$
production.
No requirements are made on the number of non-central jets.
 
The efficiency of the event preselection is 13\% (9\%) for SM ggF (VBF)
$HH$ events.
The number of events selected in data in this inclusive signal
region is 1874. With this selection, approximately 45\% of the continuum background consists of events with two genuine $b$-jets and two prompt photons, 40\% consists of events with two genuine prompt photons and at least one misidentified $b$-jet, and 15\% consists of events with at least one misidentified photon.

\subsection{Event categories}
\label{ssec:categories}
The kinematic properties of Higgs boson pair production, especially
$m_{HH}$, are significantly
affected by the values of \kl and \kvv. In particular,
ggF and VBF $HH$ production with values of \kl close to the SM
expectation lead to rather large values of $m_{HH}$, while
for \kl significantly different from one the $HH$ invariant
mass spectrum is relatively soft.
Anomalous values of $\kappa_{2V}$ also lead to events,
produced via VBF, with a large invariant mass of the Higgs boson pair.
The events are therefore classified in two regions based on
the modified four-body invariant mass $\myybb^{*} = m_{\yybb} - (m_{\bb} - 125~\GeV) -
(m_{\gamma\gamma} - 125~\GeV)$:
a \textit{high mass} ($\myybb^{*}>350$~\GeV) region and a
\textit{low mass} ($\myybb^{*}\leq 350$~\GeV) region.
The use of $\myybb^{*}$ over $\myybb$ improves the signal mass
resolution due to the cancellation of detector resolution
effects~\cite{HDBS-2018-34_custom}.
 
In each of the two $\myybb^{*}$ regions, a dedicated
boosted-decision-tree (BDT) discriminant
is trained to distinguish $HH$ signals from the background
arising from $H\to\gamma\gamma$ decays in single Higgs boson
production events and from the continuum diphoton background from
$t\bar{t}\gamma\gamma$ and $\gamma\gamma$+jets events.
The training is performed with the \AtlasMC{XGBoost}
program~\cite{Chen:2016:XST:2939672.2939785}
using only simulated event samples.
In the high mass region, the signal samples used for training include
SM ggF and VBF $HH$ events, as well
as the five non-SM samples of the VBF $HH$ basis.
In the low mass region, the signal samples consist of
non-SM ggF $HH$ events corresponding to $\kl = 10$
and $\kl = 5.6$,
plus the same five non-SM VBF $HH$ basis samples.
The choice of $\kl = 5.6$ corresponds to a large anomalous value
of $\kl$ that is not yet excluded with a high confidence level by
the previous search in this channel. However, it is observed that
the training is relatively stable for variations of the order of unity on the $\kl$ value used in training.

The BDT discriminant uses the same input variables that were used for
the analogous multivariate discriminant in Ref.~\cite{HDBS-2018-34_custom}
(denoted by \textit{baseline} variables), complemented
by a set of additional observables that provide further discrimination
between the background and the signal, mainly from VBF $HH$ production.
The baseline variables include kinematic properties of the two photon
and the two $b$-jet candidates, the scalar sum $H_\mathrm{T}$ of the
\pT of all the jets, and the magnitude \etmiss{} and direction
$\phi^\mathrm{miss}$ of the missing transverse momentum vector
$\vec{p}_\mathrm{T}^\mathrm{miss}$~\cite{PERF-2016-07}.
Another baseline variable is the \textit{single-topness} $\chi_{Wt}$,
quantifying how likely any three-jet combination in the event is to
originate from a $t\to Wb \to q\bar{q}'b$ decay:
 
\begin{equation}
\label{eq:topness}
\chi_{Wt} = \text{min}\sqrt{{ \left(\frac{m_{j_1j_2} - m_W}{m_W} \right)}^2 + { \left(\frac{m_{j_1j_2j_3} - m_t}{m_t} \right)}^2},
\end{equation}
where $m_W$ and $m_t$ are the masses of the $W$ boson and of the top quark,
and the minimum is evaluated over all combinations of any three jets in the
event, with no requirements on whether they are $b$-tagged.

The additional variables include,
for events with at least four jets, the $\pT$, $\eta$, $\phi$, and discrete $b$-tagging score of the third and fourth jets.  
Events with at least four jets can arise from VBF $HH$ production,
in which the scattered quarks responsible for the VBF process hadronise
after having radiated a weak boson and produce two forward, high-momentum
jets (`VBF jets').
In events with exactly four selected jets, the two non $b$-tagged jets
are considered as VBF-jet candidates.
In events with at least five selected jets
(about 25\% of the VBF $HH$ events passing the previous requirements
according to the simulation),
the two non $b$-tagged jets that are considered as
VBF-jet candidates are determined by means of a BDT classifier
(`VBF-jet tagger').
The inputs of the VBF-jet tagger consist of:
(i) for each non $b$-tagged jet $j$, its
$\pT$, $\eta$,
and
$\Delta\eta$ and $\Delta R$
separations from the $\gamma\gamma b\bar{b}$ system; (ii)
for each $jj$ pair, its invariant mass,
$\Delta\eta$ between the two jets,
$\Delta\eta$ and $\Delta R$ separations from the $\gamma\gamma b\bar{b}$ system,
and
$\pT$, $\eta$,
and invariant mass of
the $\gamma\gamma b\bar{b} jj$ system.
The BDT is trained on simulated SM VBF $HH$ events
using the pair of jets matched to the scattered quarks as signal,
and all other pairs of jets as background.
After training, the VBF-jet tagger is applied to all possible jet
pair combinations in data and simulated events, and the jets
belonging to the pair with the highest tagger score are considered
as VBF-jet candidates.
Their invariant mass and pseudorapidity difference are then used as
input variables for the event classification BDTs.
In simulated VBF $HH$ events with at least three non $b$-tagged jets,
the VBF-jet tagger is able to correctly identify
the VBF-jet pair in 95\% of events.
 
A second set of additional variables used as input to the
event classification BDTs consists of
event-level kinematic quantities
such as $\mbbyy^*$ and
the angular separation $\Delta R(\gamma,\gamma)$ ($\Delta R(b, \bar{b})$)
between the two photon ($b$-tagged jet) candidates.
Finally, three event-shape observables are also
used: the
transverse sphericity $S_{\perp}$~\cite{STDM-2011-33}, the planar flow
$Pf$~\cite{Almeida:2008yp}, and the transverse momentum balance, defined as
 
\begin{equation}\label{eq:pt-balance}
\pt^\text{balance} = \frac{| \vec{p}_\mathrm{T}^{\gamma_1} + \vec{p}_\mathrm{T}^{\gamma_2} + \vec{p}_\mathrm{T}^{b_1} +  \vec{p}_\mathrm{T}^{b_2}|}{|\vec{p}_\mathrm{T}^{\gamma_1}| + |\vec{p}_\mathrm{T}^{\gamma_2}| + |\vec{p}_\mathrm{T}^{b_1}| +  |\vec{p}_\mathrm{T}^{b_2}|}.
\end{equation}
 
The relative weights of the training samples, as well
as the values of the \AtlasMC{XGBoost} hyperparameters, are tuned using a
Bayesian optimisation algorithm that maximises the expected
combined number-counting significance $Z$~\cite{Cowan:2010js}
of a benchmark signal using the signal and background yields
in each category in the diphoton invariant mass
range $120~\GeV<m_{\gamma\gamma}<130$~\GeV, as described below.
 
After training, three categories (labelled `High Mass $i$', $i=1..3$) in
the high mass region and four categories (labelled `Low Mass $i$', $i=1..4$)
in the low mass region are defined based on the high mass region and
low mass region BDT discriminants, with a higher category index $i$
corresponding to higher BDT scores and more signal-like events.
Events from the inclusive signal region are thus classified in
seven orthogonal exclusive signal regions based on the value of
$\myybb^{*}$ and of the BDT scores.
Events with a BDT score lower than the threshold defining the category
with the lowest index in the corresponding low or high mass region
are discarded.
The values of the BDT scores used to define the categories
are chosen by maximising the combined number-counting significance
of all categories in a region for a benchmark signal using expected
signal and background yields in the diphoton invariant mass range
$120~\GeV<m_{\gamma\gamma}<130$~\GeV.
During this optimisation process, each category must contain at least
nine expected continuum background events in the $\myy$
\textit{sidebands}, \textit{i.e.} excluding the region
$120~\GeV<\myy<130$~\GeV, in order to have sufficient events to
constrain the shape of the diphoton invariant mass distribution of the
continuum background when the selection is evaluated on the data.
In the high mass region, the signal yield is computed from the sum of the
expected SM ggF and VBF $HH$ contributions, while in low mass region,
the signal yield is computed from the ggF $HH$ $\kl = 5.6$ and
VBF $HH$ $\kl = 10$ predictions.
 
\begin{figure}[!hbtp]
\centering
\subfloat[]{ \includegraphics[width=0.48\textwidth]{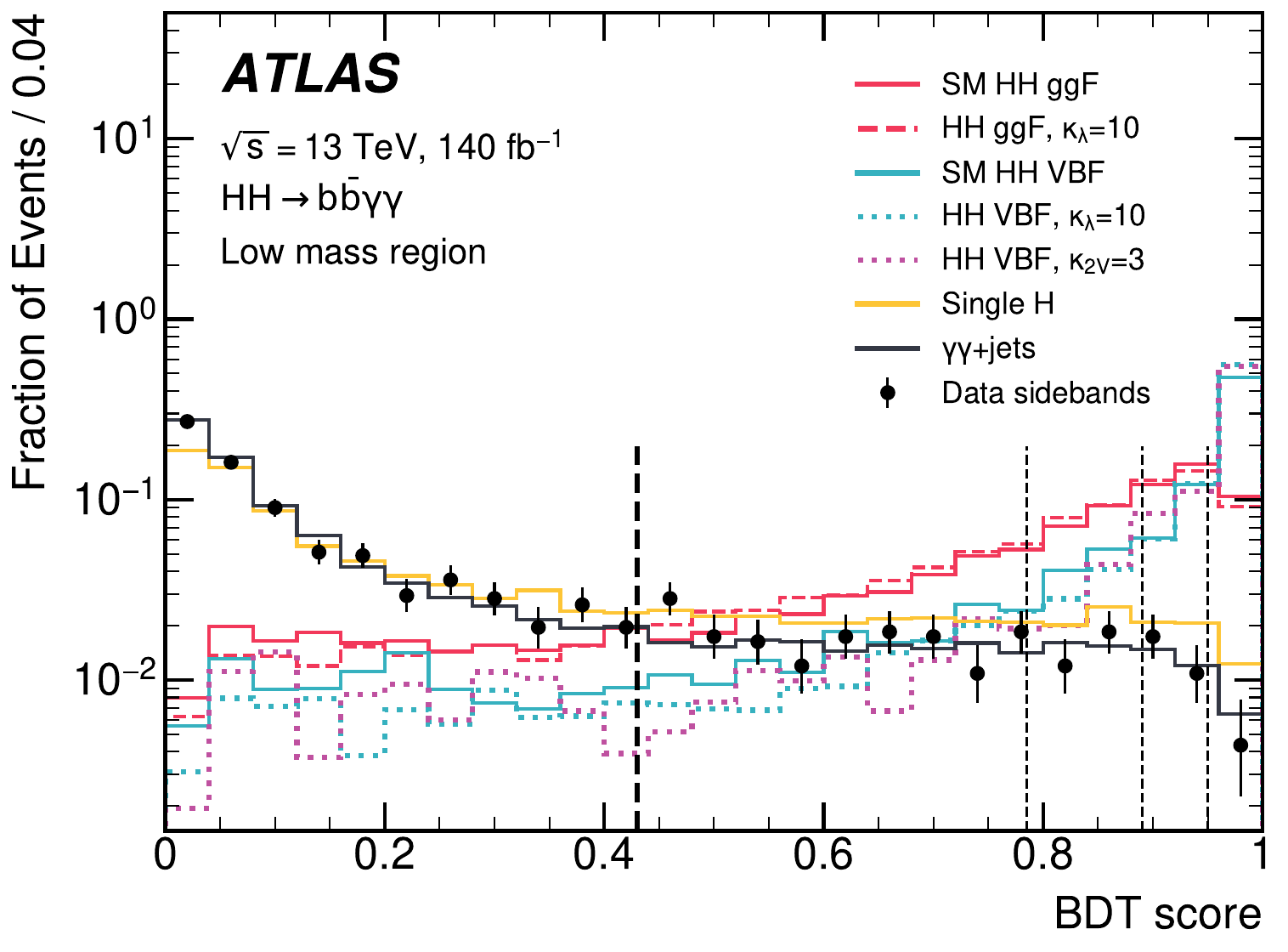}\label{fig:BDT_LM}}
\subfloat[]{ \includegraphics[width=0.48\textwidth]{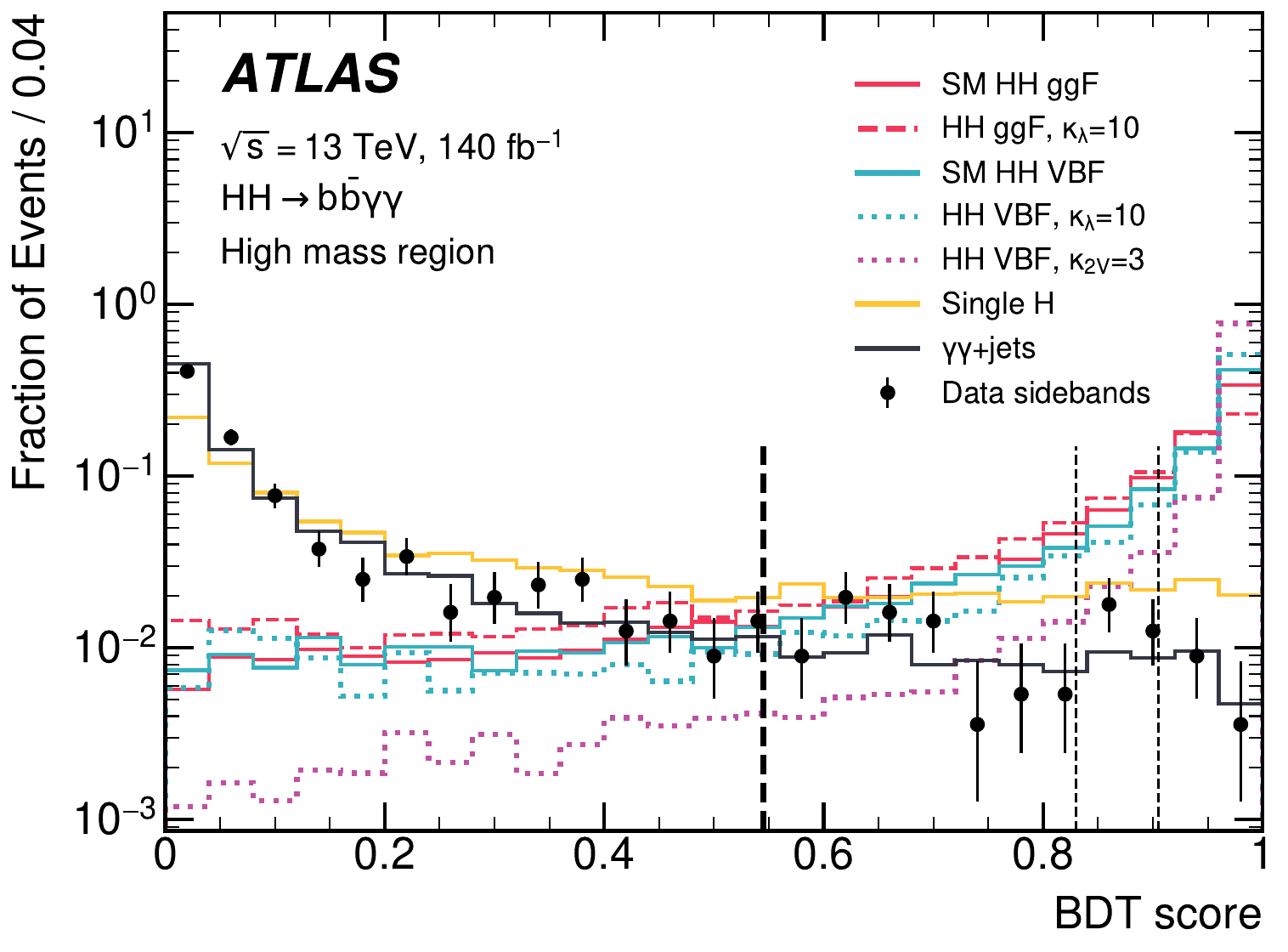}\label{fig:BDT_HM}}
\caption{
BDT score distributions for simulated ggF and VBF $HH\to\bbyy$
signal events and simulated background events from
nonresonant $\gamma\gamma$ + jets and singly produced Higgs
bosons decaying into $\gamma\gamma$ for the
\protect\subref{fig:BDT_LM} low and
\protect\subref{fig:BDT_HM} high mass regions. The data in the
$m_{\gamma\gamma}$ sidebands, which are not expected to be populated by
single nor double Higgs boson events,
are also shown compared with the $\gamma\gamma$ + jets sample.
The latter comprises the majority of the nonresonant diphoton background
and is used in the training of the BDT. All distributions are normalised to unity.
The vertical dashed lines correspond to the thresholds used to define
the event categories. Events with a BDT score between 0 and the lowest
threshold (thick dashed line) are discarded. Events satisfying the lowest
threshold are categorised as Low Mass $i$, $i = 1..4$ (High Mass $i$, $i = 1..3$),
with a higher category index $i$ corresponding to higher BDT scores and more
signal-like events.}
\label{fig:BDT_output_Low_and_High_mass}
\end{figure}

The BDT discriminant distributions in the low and high mass regions
observed in data in the \myy sidebands are shown in
Figure~\ref{fig:BDT_output_Low_and_High_mass}.
Also illustrated for comparison are the expected BDT score distributions
for the dominant nonresonant background from the $\gamma\gamma$ + jets sample,
the resonant single Higgs boson background, and
the ggF and VBF $HH$ signals for different values of
$\kl$ and $\kvv$.
The values of the BDT scores that define the categories
are represented by vertical dashed lines.
In total, 340 events in the range of $105~\GeV<\myy<160$~\GeV\
are retained from the 1874 passing the initial
preselection.
 
\section{Signal and background modelling of the diphoton mass spectrum}
\label{sec:sb_modelling}
The signal, resonant and nonresonant background yields in each category
are determined from unbinned fits to the diphoton
invariant mass distributions in the signal regions, as described in
Section~\ref{sec:results}.
The signal and background $\myy$ distributions in each category
are independently modelled by means of analytical functions
chosen as follows.
 
The $\myy$ distributions of signal events and resonant background events
from single Higgs bosons decaying into $\gamma\gamma$
are described by double-sided Crystal Ball
functions~\cite{HIGG-2013-08,dscb1}.
The shape parameters are obtained from fits to simulated SM $HH$
events, and then either fixed in the final fits (parameters
describing the tail of the distribution) or constrained around the
initial values within the uncertainties resulting from the photon energy
calibration.
The same model is found to describe selected single Higgs boson and
Higgs boson pair events well for both SM and non-SM coupling values.
Signal + background fits performed on a combination of signal and resonant
background events from simulation and the expected nonresonant background
distribution show negligible signal yield non-closure resulting from
this assumption.
 
The $\myy$ distributions of the nonresonant diphoton background
are modelled with exponential functions, whose normalisation and
shape parameters are obtained from the fit to the data.
The chosen exponential model in each category has two degrees
of freedom and is found to describe the data well in the $\myy$
sidebands, as well as the background-only template obtained with
the \Sherpa[2.2.12] $\gamma\gamma b\bar{b}$ sample normalised to the
data in the $\myy$ sidebands.
The \textit{spurious signal}~\cite{HIGG-2013-08,ATL-PHYS-PUB-2020-028}
is defined as the maximum absolute value of the bias
on the fitted signal yield in multiple signal+background fits to
the background-only template, performed with varying mass assumptions on
$m_H \in [123, 127]~\GeV$ in intervals of $0.5~\GeV$.
For each of the exponential models, the spurious signal is smaller than 20\%
of the statistical uncertainty in the expected fitted signal
yield, plus twice the statistical uncertainty in the spurious signal
itself. Alternative models with the same numbers of degrees of freedom,
such as power functions of $\myy$, performed similarly to the
exponential model.
While the nominal templates are constructed with simulated
$\gamma\gamma b\bar{b}$  events only, alternative templates accounting
for potential shape differences due to other background components
such as $\gamma\gamma q\bar{q}$ ($q \neq b$), $\gamma j$, and $jj$
do not significantly alter the spurious signal value or the quality
of the exponential fit.


\section{Systematic uncertainties}
\label{sec:systematics}
 
Systematic uncertainties affect the shape and normalisation of the
diphoton invariant mass distributions of the Higgs boson pair signal and
single Higgs boson backgrounds.
Nevertheless, due to the limited number of events and the small
signal-to-background ratio, the impact of the
systematic uncertainties is small compared with that of the statistical
uncertainties.
 
The systematic uncertainties are computed separately for the ggF and
VBF $HH$ production modes and for the various single Higgs boson
production modes. Those from the same source are correlated between processes.
For the ggF (VBF) $HH$ signal, for each
source of uncertainty the corresponding estimate
is obtained by taking the envelope of values computed using both the SM
and the $\kl=10$ simulated
event sample (using the six VBF basis simulated event samples).
 
The uncertainty in the full Run 2 integrated luminosity
is derived from dedicated measurements~\cite{DAPR-2021-01}
using the LUCID-2~\cite{LUCID2} detector.
The diphoton trigger efficiency uncertainty
is
evaluated using radiative $Z$ boson decays and with
events collected using prescaled lower-threshold
triggers~\cite{TRIG-2018-05}.
The uncertainty in the vertex selection efficiency
is
evaluated by comparing the reconstruction efficiency of
photon-pointing vertices in $Z \rightarrow e^+e^-$ events in data with
that in simulation~\cite{HIGG-2020-16_custom}.
 
The uncertainties in photon identification and isolation efficiencies
are
determined from control samples of prompt photons from
photon+jet production and from radiative $Z$ boson decays
and electrons~\cite{EGAM-2018-01}.
The uncertainties in the photon energy scale and resolution
are determined from control samples of electrons from $Z$ boson and
$J/\psi$ decays and of photons from radiative $Z$ boson
decays~\cite{EGAM-2018-01}.
 
The uncertainties in the jet energy scale and resolution are
determined from control samples of jets recoiling against
well calibrated particles such as photons, $Z$ bosons or
already calibrated jets~\cite{JETM-2018-05_custom}.
Additional uncertainties from
the simulation account for potential differences between the response
for $b$-jets and jets from gluons and light quarks.
The uncertainties in the flavour-tagging efficiencies and
misidentification rates are estimated by using \ttbar{} events for $b$-
and $c$-jets and $Z$+jets events for light-flavour
jets~\cite{FTAG-2019-07_custom,FTAG-2020-08,ATLAS:2023lwk}.
 
Theoretical uncertainties due to missing higher-order
terms in the perturbative expansion of the cross-section,
the PDF set, and the value of \alphas affect the total
expected yields of single Higgs boson and Higgs boson pair events, and
their fractional contributions to each category.
These uncertainties are evaluated
by considering alternative choices of factorisation and
renormalisation scales, PDF sets, and the value of \alphas.
For SM Higgs boson pair production, the values of the QCD scale and PDF+\alphas
total cross-section uncertainties are taken from
Ref.~\cite{DiMicco:2019ngk}.
For SM $HH$ production through ggF, the QCD scale and PDF+\alphas cross-section
uncertainties are further combined with the top-quark mass scale
uncertainty according to the prescription described in
Ref.~\cite{Baglio:2020wgt}. The uncertainties in the $H \rightarrow
\yy$ and $H \rightarrow \bb$ branching ratios are also
included~\cite{deFlorian:2016spz}.
 
For the signal and the ggF, VBF, and \ttH\ single Higgs boson processes, the
uncertainty due to the choice of parton shower model is evaluated
by comparing the predictions of the nominal simulation using the
\PYTHIA[8] model with an alternative simulation in which
the same generator-level events are showered with \HERWIG[7].
An additional 100\% uncertainty in the yields of single Higgs boson
ggF, VBF and $WH$ production modes is applied, motivated by studies of
heavy-flavour production in
association with top-quark pairs~\cite{TOPQ-2017-12,TOPQ-2012-16-corr}
and $W$ boson production in association with
$b$-jets~\cite{STDM-2012-11}.
 
For the ggF $HH$ process with $\kl \neq 1$, a systematic uncertainty is
assigned to the $\kappa_{\lambda}$ reweighting procedure by computing
for each category the maximum deviation between the expected yields
determined from the ggF $HH$ sample generated with $\kl=10$ and
the sample generated with $\kl=1$ and reweighted to $\kl=10$.
For the VBF $HH$ process, a similar uncertainty for the potential
non-closure of the procedure used to calculate the expected
yield for any value of $\kl$ and $\kvv$ from a linear
combination of the six basis samples is determined for each
category. It is calculated as the maximum difference, for the seven
validation samples described in Section~\ref{sec:datamc}, between
the expected yield calculated with the validation sample and
that obtained from the linear combination approach.

An additional uncertainty in the signal yield is due to the choice of
the background model and is assumed to be equal to the spurious signal
described in Section~\ref{sec:sb_modelling}.
The larger equivalent integrated luminosity of the \Sherpa[2.2.12]
$\gamma\gamma b\bar{b}$ sample used to create the background-only template,
compared with that of the \Sherpa[2.2.4] $\gamma\gamma+$jets sample used
in the previous search published in Ref.~\cite{HDBS-2018-34_custom},
is more effective at suppressing statistical fluctuations in the template
that would otherwise lead to overestimated spurious signals.
As a consequence, the spurious signal obtained with the background
template from the \Sherpa[2.2.12] $\gamma\gamma b\bar{b}$ sample
in each category ranges between 10\% and 50\% of that from the
background template produced with the \Sherpa[2.2.4] $\gamma\gamma+$jets
sample.
Its impact on the expected upper limit on the $HH$ signal strength $\mu_{HH}$,
defined as the ratio of the Higgs boson pair production cross-section
to its SM prediction,
is thus at the permille level, compared to 3\% in the previous analysis.

The impacts of the systematic uncertainties in the expected 95\% CL
upper limit on $\mu_{HH}$, determined with the statistical
interpretation described in the next section, are listed in
Table~\ref{tab:syst}.
 
\begin{table}[!htb]
\caption{Breakdown of the dominant systematic uncertainties in
the expected $\mu_{HH}$ upper limit at 95\% CL.
The impact of the uncertainties
corresponds to the relative variation of the expected upper limit
when re-evaluating the profile likelihood ratio after fixing the
nuisance parameter in question to its best-fit value, while all
remaining nuisance parameters remain free to float.
Only systematic uncertainties with an impact
of at least 0.1\% are shown.
}
\begin{center}
\begin{tabular}{lc}
\toprule
Systematic uncertainty source &  Relative impact [\%] \\
\midrule
Experimental \\
\midrule
Photon energy resolution & 0.4 \\
Photon energy scale      & 0.1 \\
Flavour tagging          & 0.1 \\
\midrule
Theoretical \\
\midrule
Factorisation and renormalisation scale & 4.8 \\
$\mathcal{B}$($H\rightarrow \gamma\gamma, b\bar{b}$) &  0.2 \\
Parton showering model  & 0.2 \\
Heavy-flavour content   & 0.1 \\
\midrule
Background model (spurious signal) & 0.1\\
\bottomrule
\end{tabular}
\label{tab:syst}
\end{center}

\end{table}


\section{Results}
\label{sec:results}
 
The results are derived using the statistical procedures outlined
in Refs.~\cite{HIGG-2015-07, HDBS-2018-34_custom, HDBS-2022-03_custom}
from the global likelihood function
$L(\boldsymbol{\alpha},\boldsymbol{\theta})$.
The set $\boldsymbol{\alpha}$ contains the parameters of interest (POI)
of the measurement, while $\boldsymbol{\theta}$ is the ensemble of
nuisance parameters, corresponding to systematic uncertainties
constrained by auxiliary measurements
in control regions or by theoretical predictions, or to parameters
such as the continuum background yields that are {\em a priori}
unconstrained.
The function $L(\boldsymbol{\alpha},\boldsymbol{\theta})$ is the product of the
likelihood functions in each of the seven orthogonal categories,
and of constraint terms for the nuisance parameters that are not freely
floating in the fit.
For each category, the likelihood function is
determined from the corresponding signal and background
models of the $m_{\gamma\gamma}$ probability density functions
described in Section~\ref{sec:sb_modelling}, the signal and background yield
expectations for given values of $\boldsymbol{\alpha}$ and
$\boldsymbol{\theta}$,
and the observed $m_{\gamma\gamma}$ distribution in data.
 
The constraints on the coupling strength parameters,
expressed as 68\% and  95\% CL intervals, are determined
with the same procedure as that of Ref.~\cite{HDBS-2022-03_custom},
using a profile-likelihood-ratio
test statistic $\Lambda(\boldsymbol{\alpha},\boldsymbol{\theta})$
computed from the likelihood function
in the asymptotic approximation~\cite{Cowan:2010js},
where the POIs in $\boldsymbol{\alpha}$
are the coupling strength modifiers $\kappa$.
Signal strength upper limits are derived as in
Ref.~\cite{HDBS-2018-34_custom} using the CL$_\text{s}$
approach~\cite{Read:2002hq_custom} from a separate test
statistic $\tilde{q}_\alpha$ that evaluates to zero when the parameter
of interest $\alpha=\mu_{HH}$
corresponding to the cross-section under study
is lower than its maximum likelihood estimate (MLE) $\hat\mu_{HH}$.
The allowed $\kl$ interval
published in Ref.~\cite{HDBS-2018-34_custom} was determined
in a different way, from the range of $\kl$ values for which the predicted
$HH$ cross-sections are lower than the observed upper limits.
The expected results are obtained with Asimov datasets~\cite{Cowan:2010js}
generated from the likelihood function after setting all nuisance
parameters to their MLE in the fit to the data and fixing the POIs
to the values corresponding to the hypothesis under test.
The asymptotic results are found to agree within 10\%
with values obtained using pseudo-experiments.
 
\begin{figure}[htbp]
\centering
\subfloat[Low Mass 1]{\includegraphics[width=0.4\textwidth]{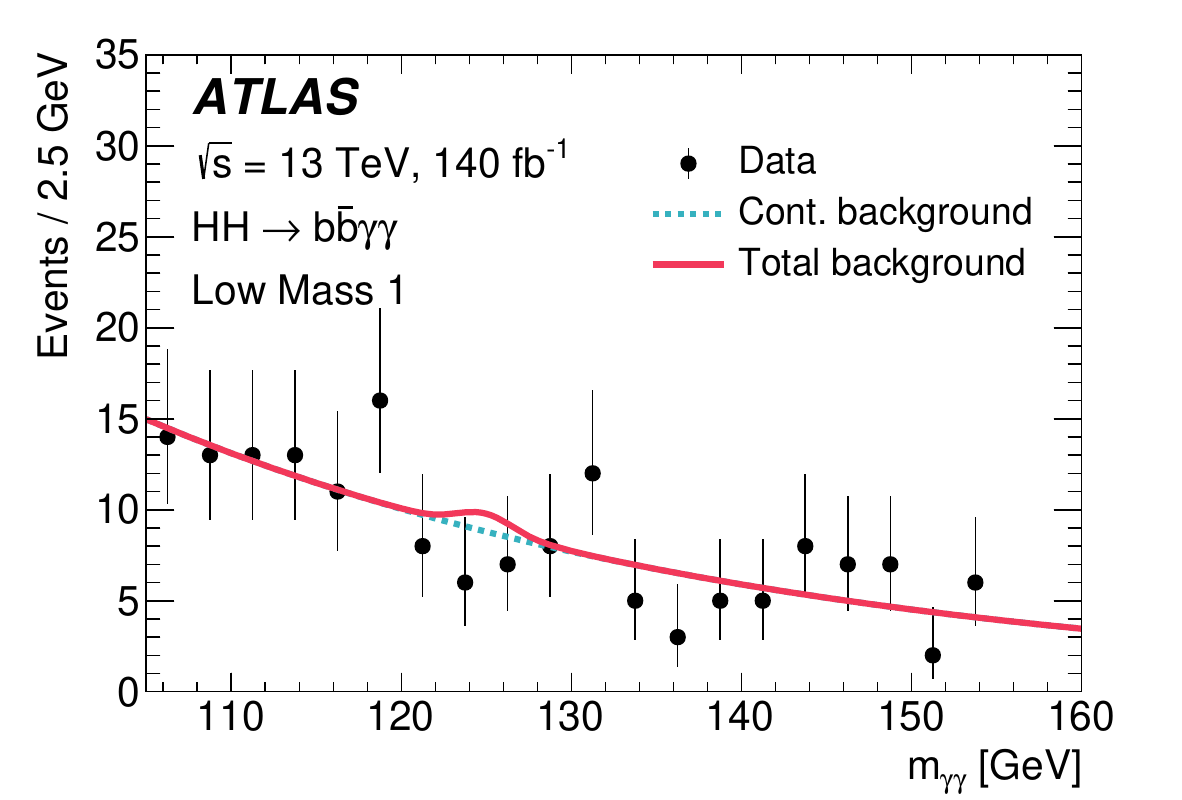}}
\subfloat[Low Mass 2]{\includegraphics[width=0.4\textwidth]{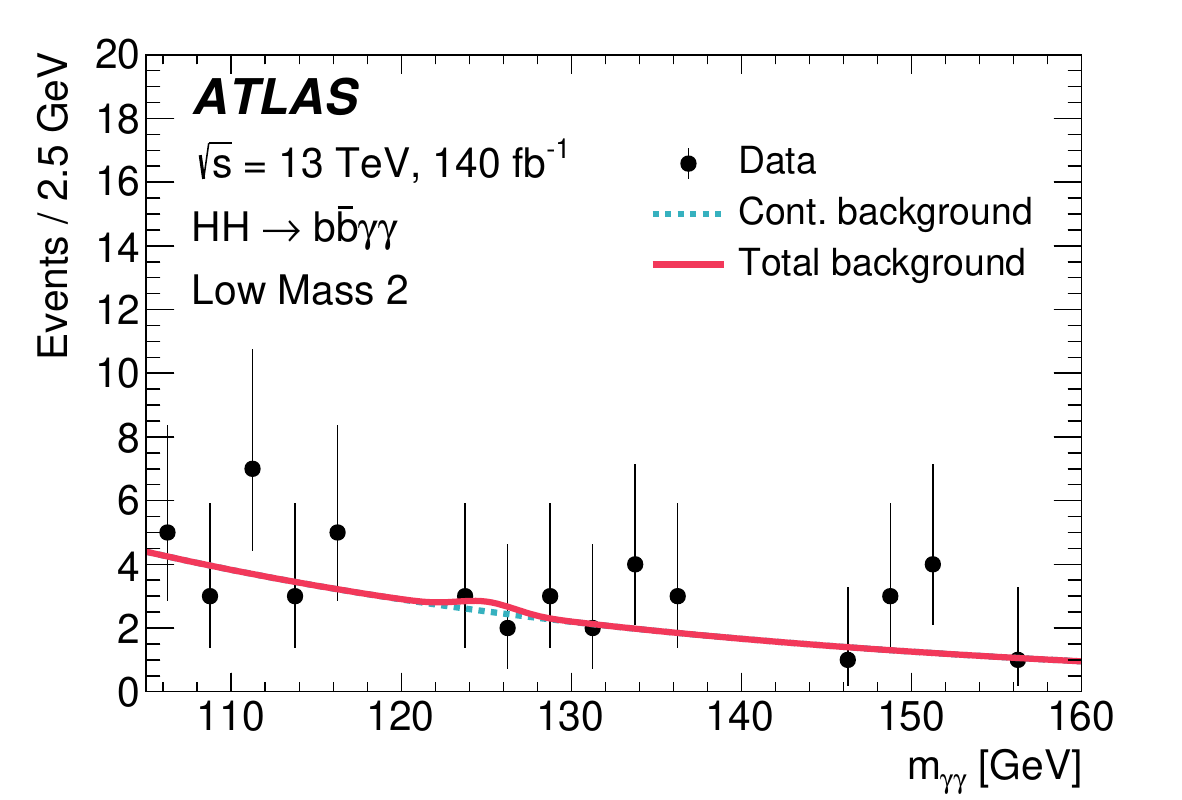}}\\
\subfloat[Low Mass 3]{\includegraphics[width=0.4\textwidth]{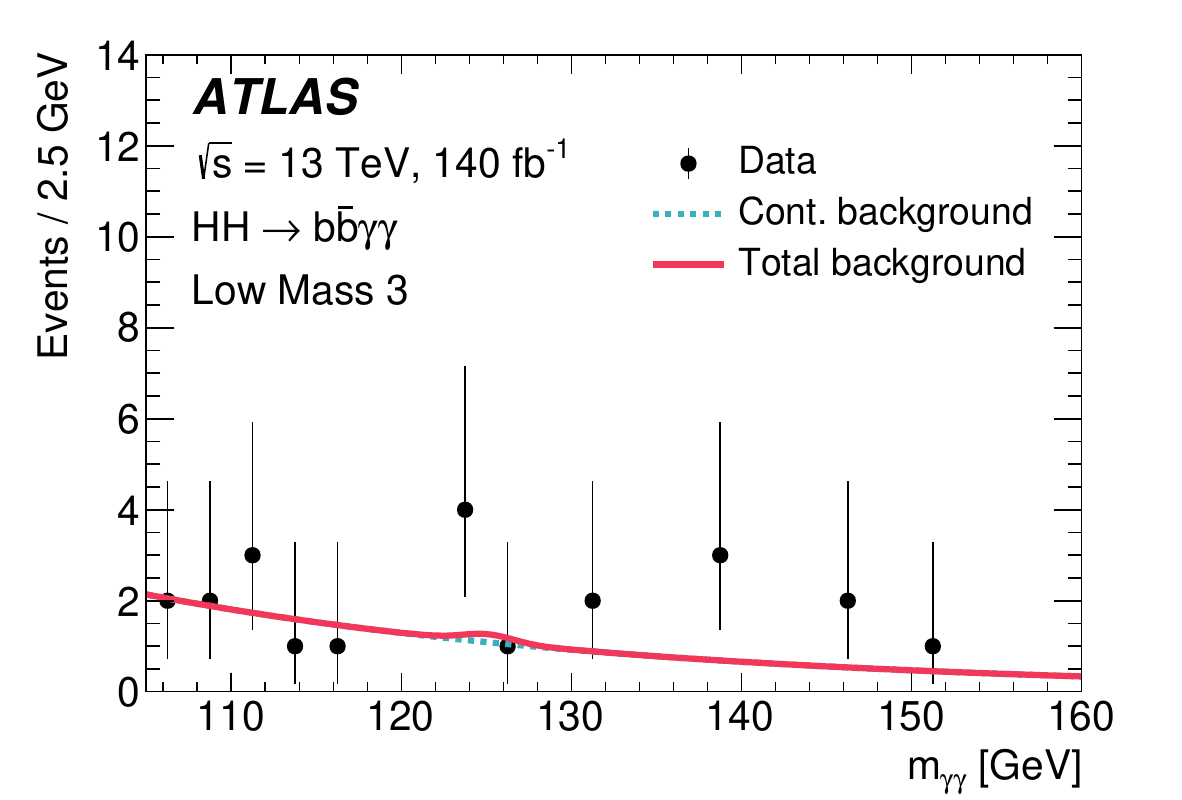}}
\subfloat[Low Mass 4]{\includegraphics[width=0.4\textwidth]{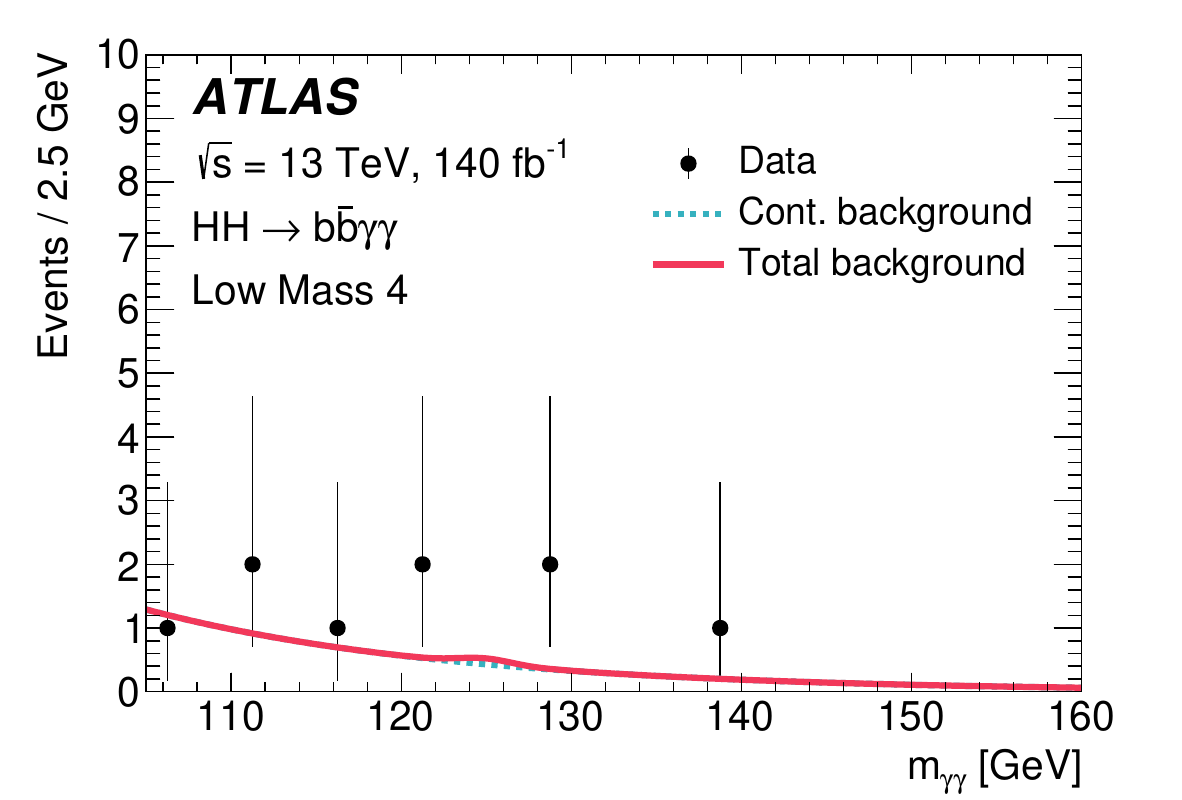}}\\
\subfloat[High Mass 1]{\includegraphics[width=0.4\textwidth]{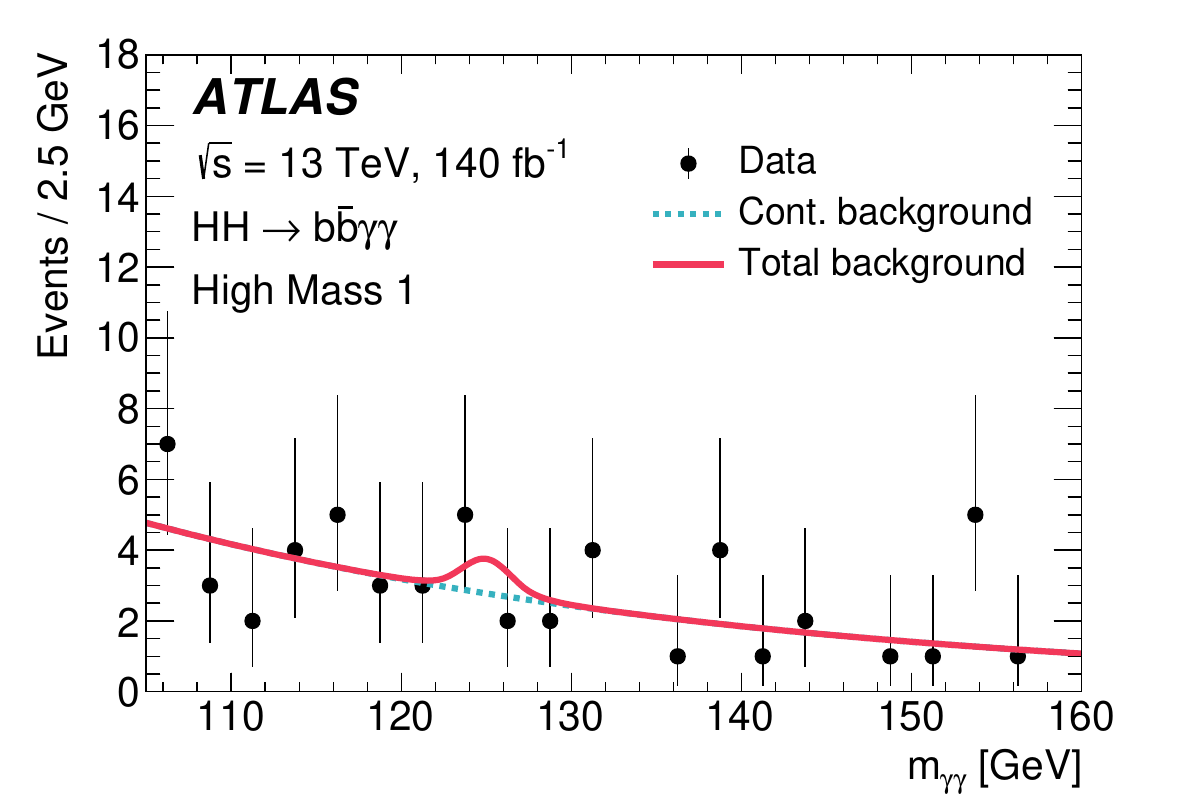}}
\subfloat[High Mass 2]{\includegraphics[width=0.4\textwidth]{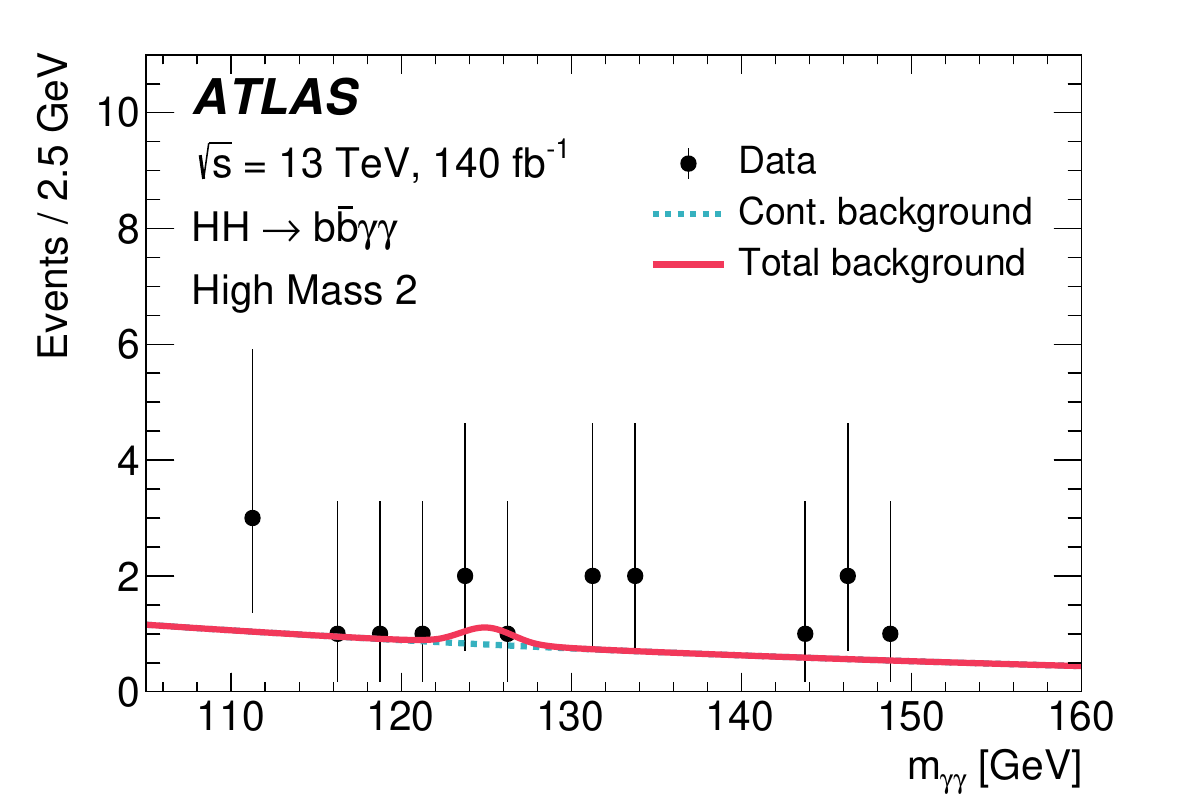}}\\
\subfloat[High Mass 3]{\includegraphics[width=0.4\textwidth]{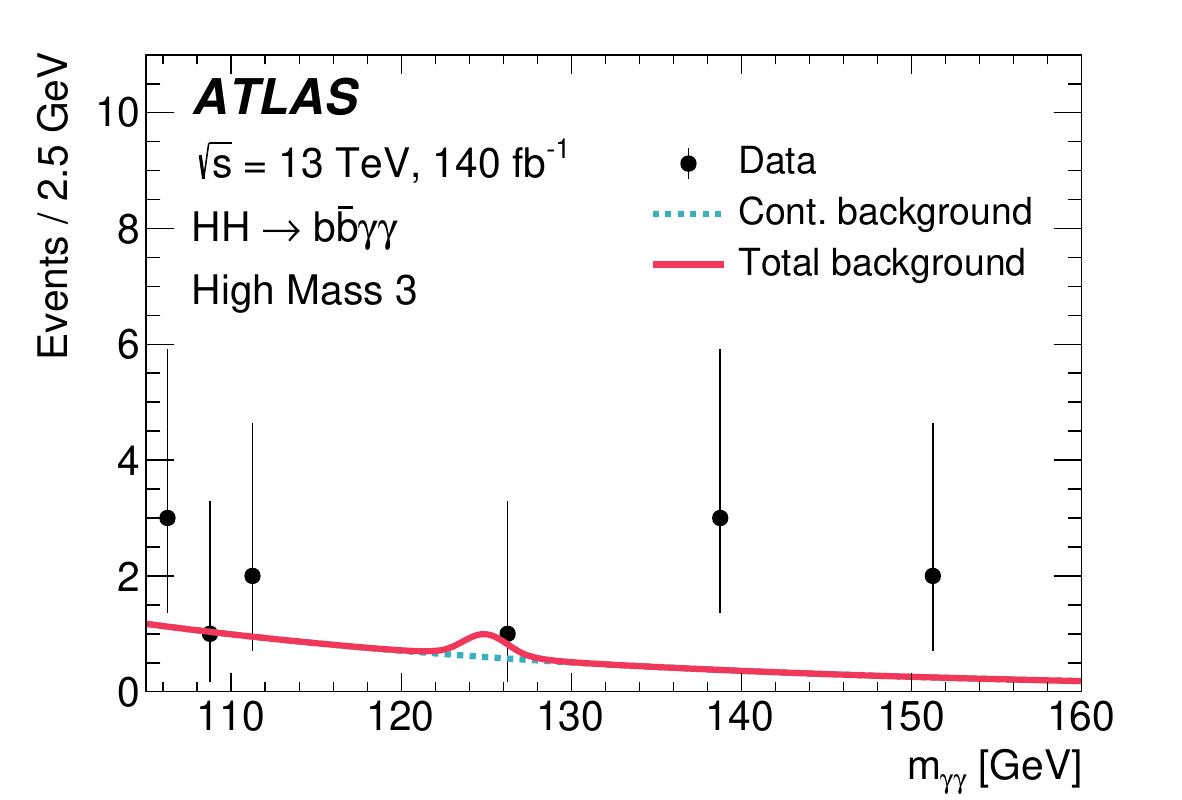}}
\caption{Comparison between the diphoton invariant mass
distribution in data (points with error bars) and the
background-only fit (solid line) for the four low mass (a--d) and
three high mass (e--g) categories of the $HH\to\bbyy$ search. In
each of the regions, a higher category
index corresponds to higher BDT scores and more signal-like
events. The solid line peaks near 125~\GeV are due to single Higgs
boson production.
}
\label{fig:sideband_fit}
\end{figure}

Figure~\ref{fig:sideband_fit} shows the result of a background-only
fit to the data, using the likelihood function $L$ after fixing the
parameters of interest corresponding to setting the signal cross-sections to zero.
Table~\ref{tab:yields} compares the number of events in the observed
data to the expected values in each category.
No significant excess over the expected background is found,
and a 95\% CL upper limit of 4.0 on the total $HH$ production
signal strength $\mu_{HH}$ (where only ggF and VBF processes
are considered) is set,
to be compared with an expected limit of 5.0 (6.4)
in the background-only $\mu_{HH}=0$ (SM $\mu_{HH}=1$) hypothesis.
If the VBF (ggF) $HH$ signal strength is fixed to the SM prediction,
the observed upper limit on the ggF (VBF) $HH$ signal strength is 4.1
(96), while the expected upper limit, computed assuming
$\mu_{\text{ggF}} = 0$ ($\mu_{\text{VBF}} = 0$), is 5.3 (145).
The observed limits are tighter than the expected ones due
to deficits in the signal regions of the most sensitive categories, as
shown in Table~\ref{tab:yields}. The compatibility between the
best-fit value of $\mu_{HH}$ and the SM expectation is approximately
1.3 standard deviations.
 
\begin{table*}[htb]
\centering
\caption{
The expected number of events (estimated by using simulation) from $HH$
signals with various $\klambda$ and $\kvv$ hypotheses and single
Higgs boson production, and the expected number of events
from the continuum background, evaluated in the $120~\GeV<\myy<130~\GeV$
window. For comparison, the number of observed data events is also shown.
The uncertainties in the $HH$ signals and single Higgs boson backgrounds
include the systematic uncertainties discussed in
Section~\ref{sec:systematics}.
Asymmetric uncertainties arise primarily from the theory calculation
of the SM ggF $HH$ cross-section and the large uncertainty in the
yield of single Higgs bosons produced in ggF events
in association with heavy-flavour jets, parameterised by a
lognormal distribution.
The uncertainty in the continuum background is given by the sum in
quadrature of the statistical uncertainty from the fit to the data
and the spurious signal uncertainty.
}
\label{tab:yields}
\resizebox{\textwidth}{!}{
\begin{tabular}{lccccccc}

\toprule
&   High Mass 1     & High Mass 2     & High Mass 3     & Low Mass 1     & Low Mass 2    & Low Mass 3     & Low Mass 4 \\
\midrule
\midrule
SM $HH (\klambda=1)$ signal         & $0.26_{-0.04}^{+0.03}$ & $0.194_{-0.032}^{+0.021}$ & $0.84_{-0.14}^{+0.10}$ & $0.048_{-0.008}^{+0.007}$ & $0.038_{-0.006}^{+0.004}$ & $0.039_{-0.006}^{+0.004}$ & $0.032_{-0.004}^{+0.004}$\\[5pt]
\phantom{Single Higgs}\ggF          & $0.25_{-0.04}^{+0.03}$ & $0.188_{-0.032}^{+0.021}$ & $0.81_{-0.14}^{+0.10}$ & $0.046_{-0.008}^{+0.007}$ & $0.036_{-0.006}^{+0.004}$ & $0.037_{-0.006}^{+0.004}$ & $0.025_{-0.004}^{+0.004}$\\[5pt]
\phantom{Single Higgs}VBF $[10^{-3}]$           & $7.9_{-0.5}^{+0.6}$ & $5.3_{-0.4}^{+0.5}$ & $29_{-3}^{+4}$ & $1.98_{-0.24}^{+0.28}$ & $1.71_{-0.14}^{+0.16}$ & $1.96_{-0.19}^{+0.21}$ & $7.4_{-0.5}^{+0.6}$\\[5pt]
\midrule
Alternative $HH (\klambda=10)$ signal  & $2.5_{-0.3}^{+0.4}$ & $1.81_{-0.20}^{+0.25}$ & $6.2_{-0.6}^{+0.8}$ & $5.0_{-0.9}^{+1.2}$ & $3.8_{-0.5}^{+0.7}$ & $3.7_{-0.6}^{+0.7}$ & $3.6_{-0.4}^{+0.4}$\\[5pt]
\phantom{Single Higgs}ggF & $2.3_{-0.3}^{+0.4}$ & $1.64_{-0.19}^{+0.25}$ & $4.9_{-0.6}^{+0.8}$ & $4.7_{-0.8}^{+1.0}$ & $3.6_{-0.6}^{+0.7}$ & $3.3_{-0.5}^{+0.7}$ & $2.04_{-0.27}^{+0.34}$\\ [5pt]
\phantom{Single Higgs}VBF & $0.231_{-0.017}^{+0.019}$ & $0.170_{-0.017}^{+0.019}$ & $1.29_{-0.14}^{+0.15}$ & $0.28_{-0.11}^{+0.20}$ & $0.23_{-0.12}^{+0.23}$ & $0.36_{-0.08}^{+0.10}$ & $1.57_{-0.16}^{+0.17}$\\ [5pt]
\midrule
Alternative VBF $HH (\kvv=3)$ signal  & $0.23_{-0.04}^{+0.04}$ & $0.20_{-0.04}^{+0.05}$ & $3.8_{-0.6}^{+0.7}$ & $0.03_{-0.02}^{+0.04}$ & $0.03_{-0.02}^{+0.06}$ & $0.048_{-0.015}^{+0.023}$ & $0.17_{-0.03}^{+0.04}$\\ [5pt]
 
\midrule
\midrule
 
Single Higgs boson background & $1.5_{-0.3}^{+0.5}$ & $0.48_{-0.10}^{+0.21}$ & $0.57_{-0.14}^{+0.25}$ & $1.72_{-0.19}^{+0.31}$ & $0.53_{-0.06}^{+0.08}$ & $0.29_{-0.07}^{+0.14}$ & $0.16_{-0.03}^{+0.06}$ \\[5pt]
 
\phantom{Single Higgs}ggF & $0.5_{-0.2}^{+0.5}$ & $0.14_{-0.09}^{+0.21}$ & $0.25_{-0.12}^{+0.25}$ & $0.29_{-0.15}^{+0.31}$ & $0.08_{-0.04}^{+0.08}$ & $0.07_{-0.06}^{+0.13}$ & $0.04_{-0.03}^{+0.06}$ \\[5pt]
\phantom{Single Higgs}$t\bar{t}H$ & $0.302_{-0.032}^{+0.034}$ & $0.069_{-0.008}^{+0.009}$ & $0.063_{-0.007}^{+0.008}$ & $0.77_{-0.08}^{+0.09}$ & $0.214_{-0.026}^{+0.029}$ & $0.100_{-0.012}^{+0.012}$ & $0.048_{-0.005}^{+0.005}$\\[5pt]
\phantom{Single Higgs}$ZH$ & $0.61_{-0.05}^{+0.06}$ & $0.174_{-0.016}^{+0.020}$ & $0.188_{-0.029}^{+0.035}$ & $0.49_{-0.04}^{+0.05}$ & $0.149_{-0.025}^{+0.028}$ & $0.069_{-0.023}^{+0.033}$ & $0.028_{-0.007}^{+0.010}$\\[5pt]
\phantom{Single Higgs}Rest & $0.17_{-0.04}^{+0.08}$ & $0.089_{-0.016}^{+0.030}$ & $0.07_{-0.02}^{+0.04}$ & $0.181_{-0.019}^{+0.030}$ & $0.089_{-0.009}^{+0.016}$ & $0.046_{-0.004}^{+0.007}$ & $0.039_{-0.004}^{+0.008}$\\[5pt]
Continuum background  & $11.3_{-1.6}^{+1.5}$ & $3.2_{-0.8}^{+0.8}$ & $2.8_{-0.8}^{+0.8}$ & $37.2_{-2.9}^{+2.9}$ & $10.8_{-1.5}^{+1.5}$ & $4.4_{-1.0}^{+0.9}$ & $1.1_{-0.5}^{+0.5}$ \\[5pt]
\midrule
Total background & $12.8_{-1.6}^{+1.6}$ & $3.7_{-0.8}^{+0.9}$ & $3.4_{-0.8}^{+0.8}$ & $38.9_{-2.9}^{+2.9}$ & $11.3_{-1.5}^{+1.5}$ & $4.7_{-1.0}^{+0.9}$ & $1.3_{-0.5}^{+0.5}$ \\[5pt]
\midrule
\midrule
Data & $12$ & $4$ & $1$ & $29$ & $8$ & $5$ & $4$\\[5pt]
\bottomrule

\end{tabular}
}
\end{table*}
 
\begin{figure}[hbt]
\centering
\subfloat[]{
\includegraphics[width=0.48\textwidth]{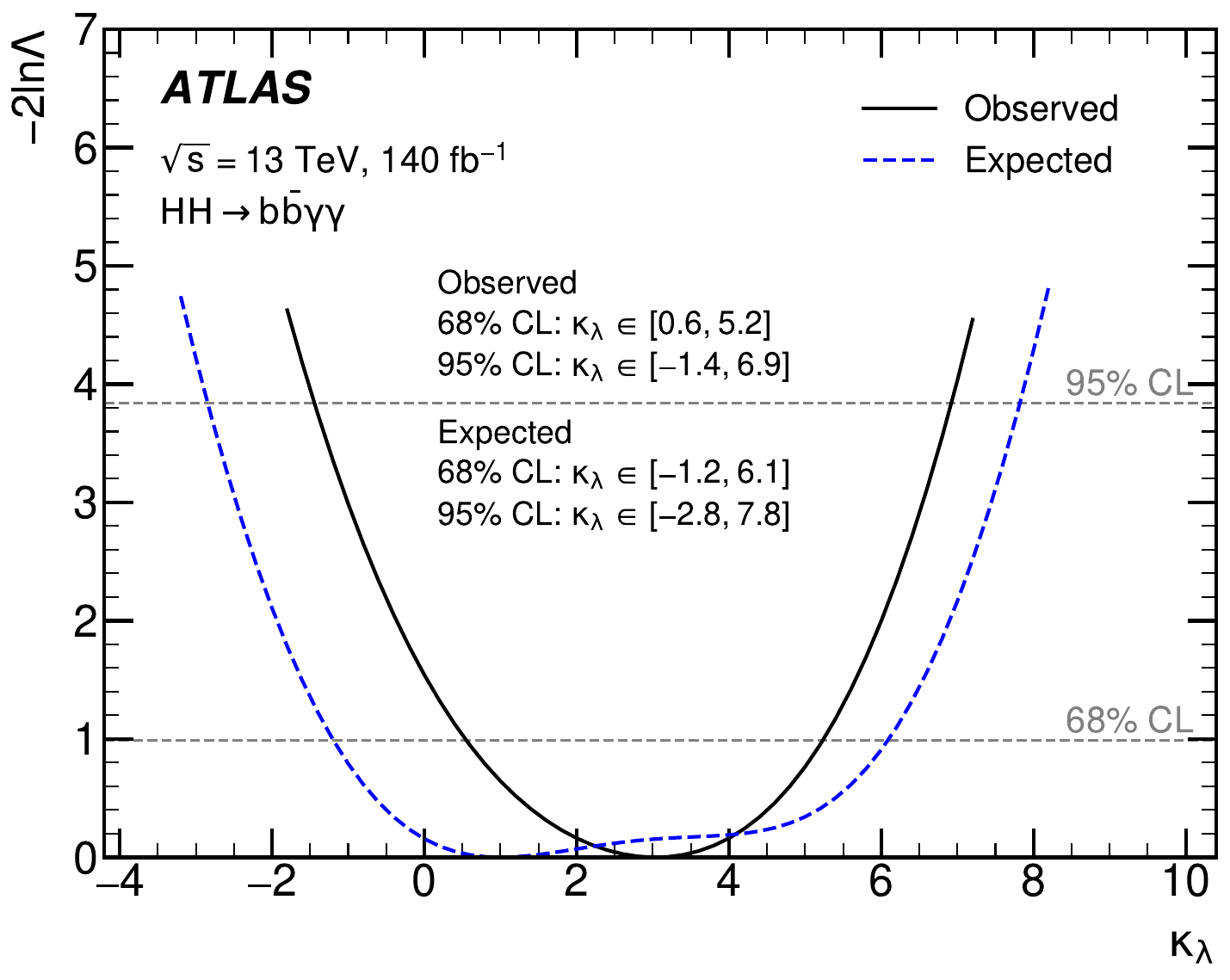}
}
\subfloat[]{
\includegraphics[width=0.48\textwidth]{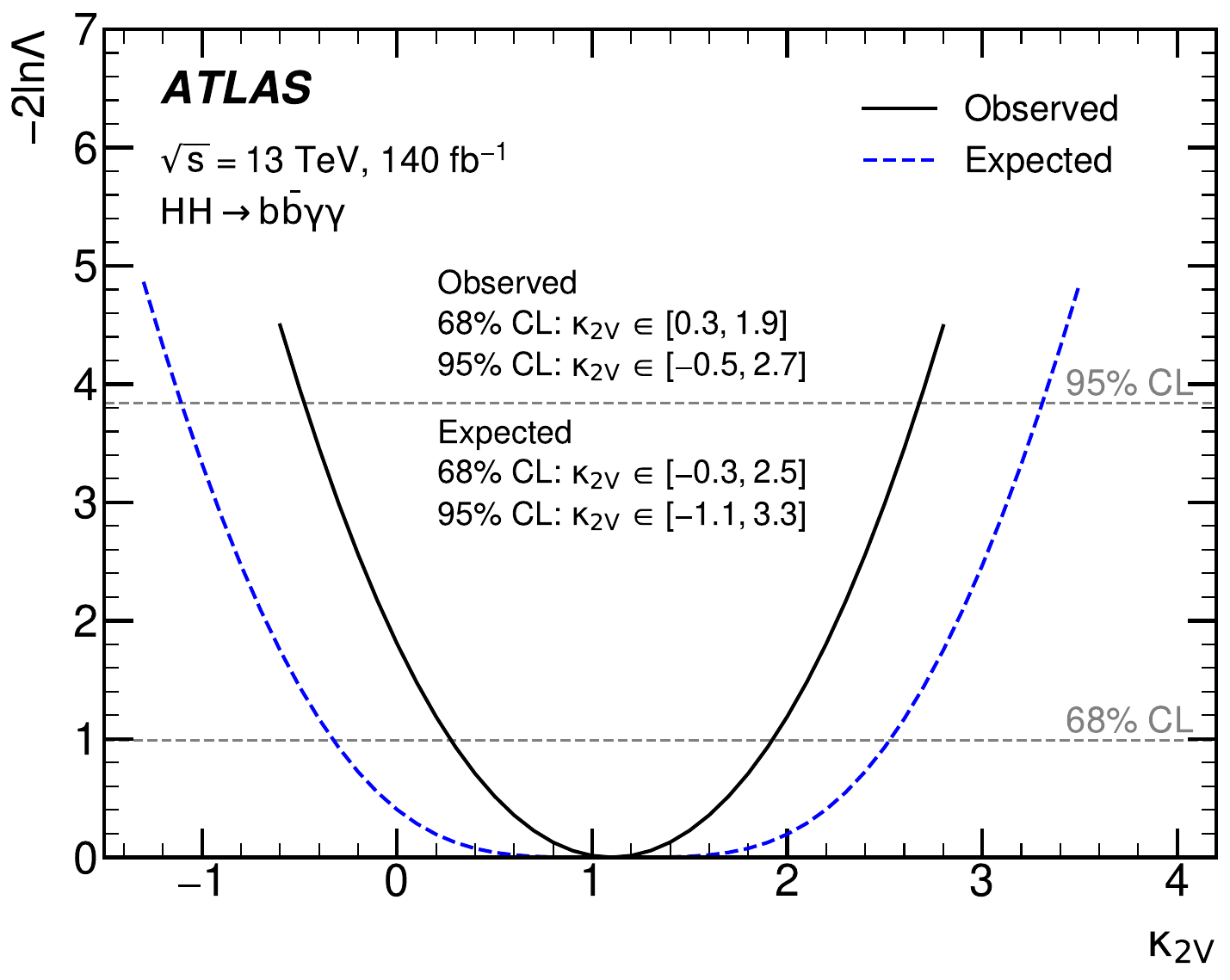}
}
\caption{Observed (solid line) and expected (dashed line) value of
$-2\ln\Lambda$ as a function of (a) $\kappa_{\lambda}$ and
(b) $\kappa_{2V}$, when all other coupling modifiers (including,
respectively, \kvv or \kl) are fixed to their SM predictions.}
\label{fig:kl_and_k2v_1d_likelihood_scans}
\end{figure}

\begin{figure}[htbp]
\centering
\includegraphics[width=0.48\textwidth]{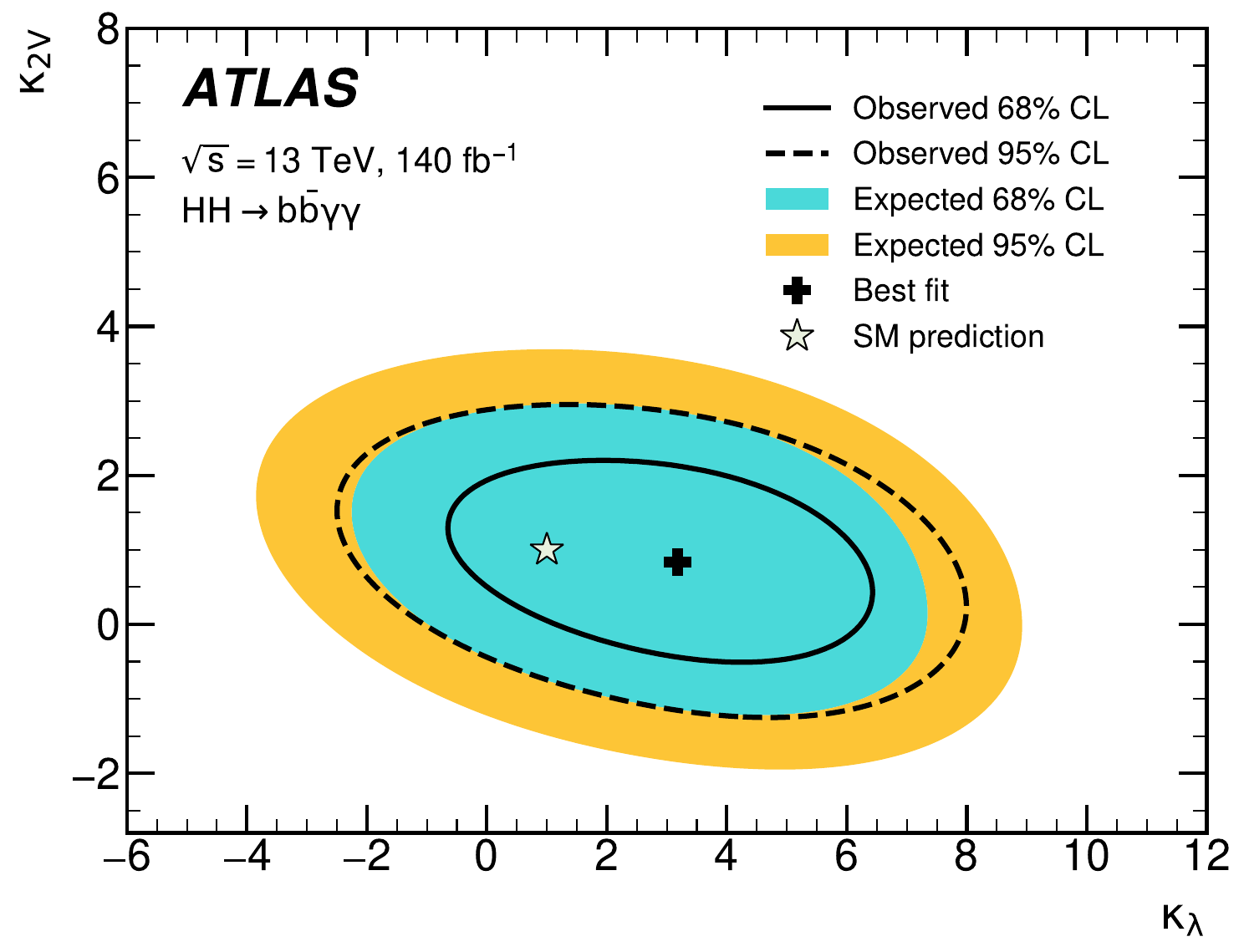}
\caption{Likelihood contours at 68\% (solid line) and 95\% (dashed line)
CL in the $(\kl, \kvv)$ parameter space, when
all other coupling modifiers are fixed to their SM predictions.
The corresponding expected contours are shown by the
inner and outer shaded regions
The SM prediction is indicated by the star, while
the best-fit value is denoted by the cross.}
\label{fig:klambda_k2v_2d_likelihood_scan}
\end{figure}

The values of $-2\ln\Lambda$ as a function of the coupling strength
factor \kl or \kvv
under the hypothesis that all other coupling modifiers are
equal to their SM predictions are shown in
Figure~\ref{fig:kl_and_k2v_1d_likelihood_scans}.
The observed (expected) constraints under this hypothesis are
$-1.4 < \kl < 6.9$ ($-2.8 < \kl < 7.8$)
and
$-0.5 < \kvv < 2.7$ ($-1.1 < \kvv < 3.3$)
at 95\% CL.
Two-dimensional constraints at 68\% and 95\% CL in the $(\kl, \kvv)$
plane are also shown in Figure~\ref{fig:klambda_k2v_2d_likelihood_scan},
when all the other coupling modifiers are fixed to their SM predictions.
 
The impact of the systematic uncertainties on the results
is small, leading to an increase of the upper limits
on the signal strengths by 6\%--7\% and to a widening of
95\% CL confidence intervals for the coupling modifiers
by 2\%--3\% relative to the case in which systematic
uncertainties are neglected.
 
Compared to the previous analysis of
Ref.~\cite{HDBS-2018-34_custom,HDBS-2022-03_custom},
the new event classification procedure leads to a
reduction in the expected upper limit on $\mu_{HH}$
by 12\% and
a reduction in the width of
the expected one-dimensional confidence interval
for $\kl$ ($\kvv$) by 6\% (17\%),
based on a consistent statistical procedure for evaluating
the 95\% confidence interval as described at the beginning of this section.
The observed upper limit on $\mu_{HH}$ is reduced by 5\%, while the observed
one-dimensional confidence interval for $\kl$ ($\kvv$) is increased by
5\% (reduced by 16\%).

The increase in the width of the observed $\kl$ confidence
interval arises from the fact that this new analysis favours larger,
less negative values of the signal strength, corresponding
to larger magnitudes of the coupling strength modifier $\kl$.
The compatibility, considering only statistical uncertainties, between
the allowed $\kl$ interval at 95\% CL
from this study and that
of Ref.~\cite{HDBS-2018-34_custom} is evaluated using a bootstrap
technique~\cite{PhysRevD.39.274}, based on the data events
passing the selection of either the previous analysis, or that
of the current one, or both.
The compatibility between the two results
is at the level of 0.3 standard deviations.


\section{Effective field theory interpretation}
\label{sec:EFT}
 
Anomalous Higgs boson self-interactions or interactions with
the other gauge fields and fermions can alter the Higgs boson
pair production cross-section and kinematics, as well as
the Higgs boson decay rates.
The results of the previous section are thus interpreted in
the context of two effective field theories to set constraints on the
Wilson coefficients of the operators describing these anomalous interactions.
 
The approach used here follows closely that described in
Ref.~\cite{ATLAS4b}, probing a similar set of operators and benchmark
points that directly affect $HH$ production:
three Wilson coefficients ($c_{hhh}$, $c_{tthh}$, $c_{gghh}$) and
seven benchmark points~\cite{Alasfar:2023xpc} of the Higgs effective
field theory, and two Wilson coefficients
($c_{H}$, $c_{H\boxed{}}$) of the $(H^\dagger H)^3$ and $(H^\dagger
H)\Box(H^\dagger H)$ operators of the `Warsaw'
basis~\cite{Grzadkowski:2010es} of the SM effective field
theory.
In the SMEFT Lagrangian, the operators $\mathcal{O}_i$ are multiplied
by coefficients $c_i/\Lambda^2$, where $\Lambda$ is the energy scale
that bounds from above the range of validity of the EFT approach.
In this study, a value of $\Lambda = 1$~\TeV\ is assumed.
In the HEFT interpretation, the only considered coefficient affecting
VBF $HH$ production is $c_{hhh}$, and thus this production mode is
always subdominant relative to ggF $HH$.
In the SMEFT interpretation, the effects of the operators on VBF $HH$
production are similarly expected to be small, since
the SMEFT preserves the Higgs doublet structure of the SM and the
corresponding cancellation between the $VVH$ and $VVHH$ diagrams
involved in VBF $HH$ production.
Consequently both interpretations consider only ggF $HH$ production
while VBF $HH$ is assumed to be negligible.
 
Predictions for ggF $HH$ production for various values of the Wilson
coefficients under study are obtained by applying an event reweighting
technique to the SM ggF $HH$ sample, similar to the method described
in Section~\ref{sec:datamc} to emulate samples with anomalous values
of \kl.
The reweighting functions are based on the particle-level $m_{HH}$
distributions predicted at NLO accuracy in the strong coupling
constant for alternative values of the EFT coefficients.
For the HEFT interpretation, the functions are taken directly from
Ref.~\cite{Gröber:2843280}, while for the SMEFT intepretation, they
are computed using \POWHEGBOX[v2] with the SMEFT@NLO
model~\cite{Degrande:2020evl}.
For the SMEFT interpretation, a similar reweighting function is also
derived for single Higgs boson processes, but instead using the
differential distribution of the Higgs boson transverse momentum.
 
Uncertainties related to PDF, $\alpha_s$, and missing higher-order
terms in the prediction are included by taking for each analysis
category the envelope of the uncertainties from each source,
determined with the same procedure as that described in
Section~\ref{sec:systematics}.
In addition, a non-closure uncertainty is estimated by comparing the
expected yields from dedicated samples corresponding to specific
values of the anomalous couplings to those from the reweighting
procedure described above in categories reproducing the analysis
selections at generator level.
These uncertainties in the expected yield are generally of the order of 10\% or less in each category and have a small impact on the
results.
 
\begin{figure}[htbp]
\centering
\subfloat[]{\includegraphics[width=0.45\textwidth]{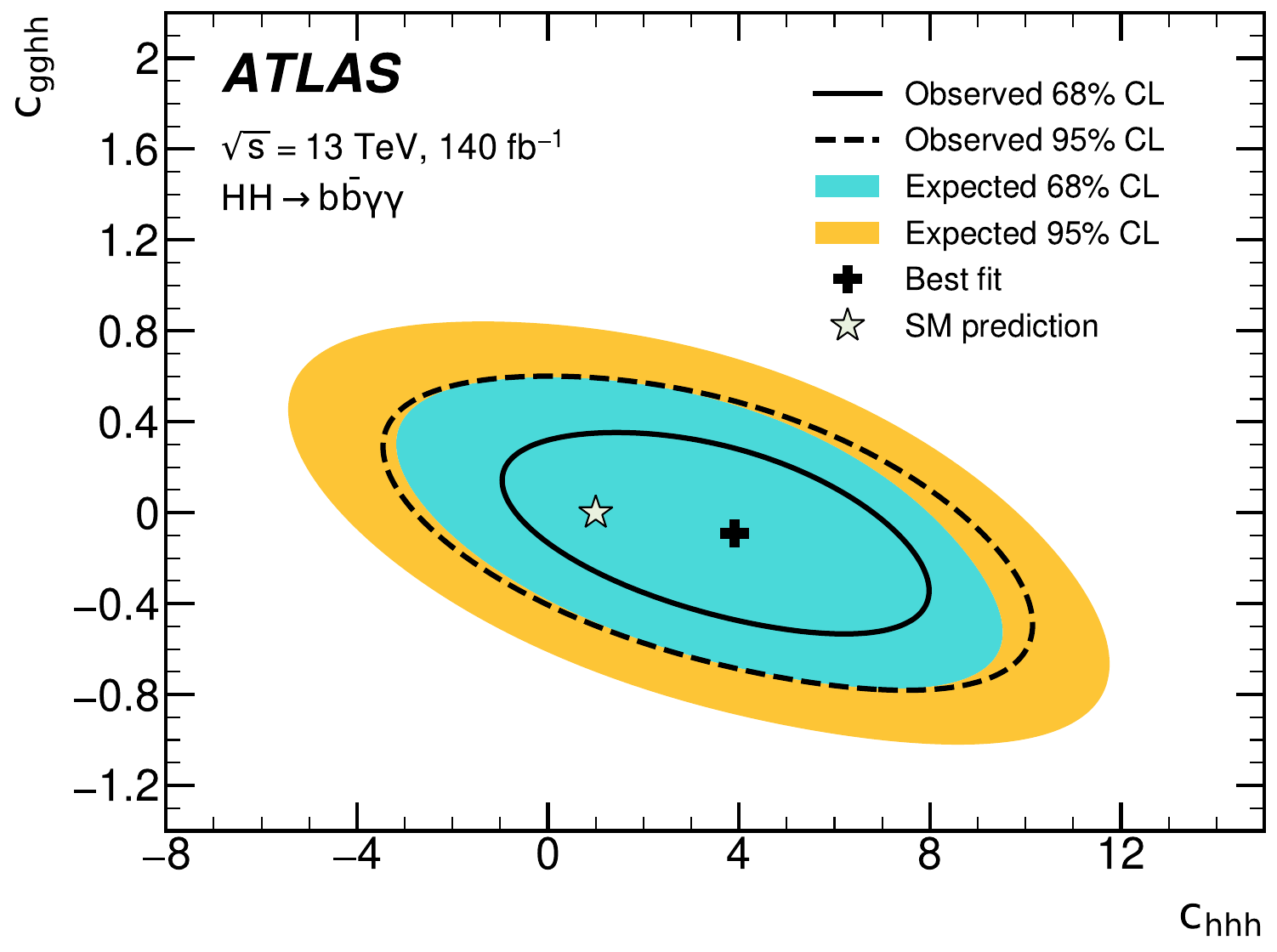}}
\subfloat[]{\includegraphics[width=0.45\textwidth]{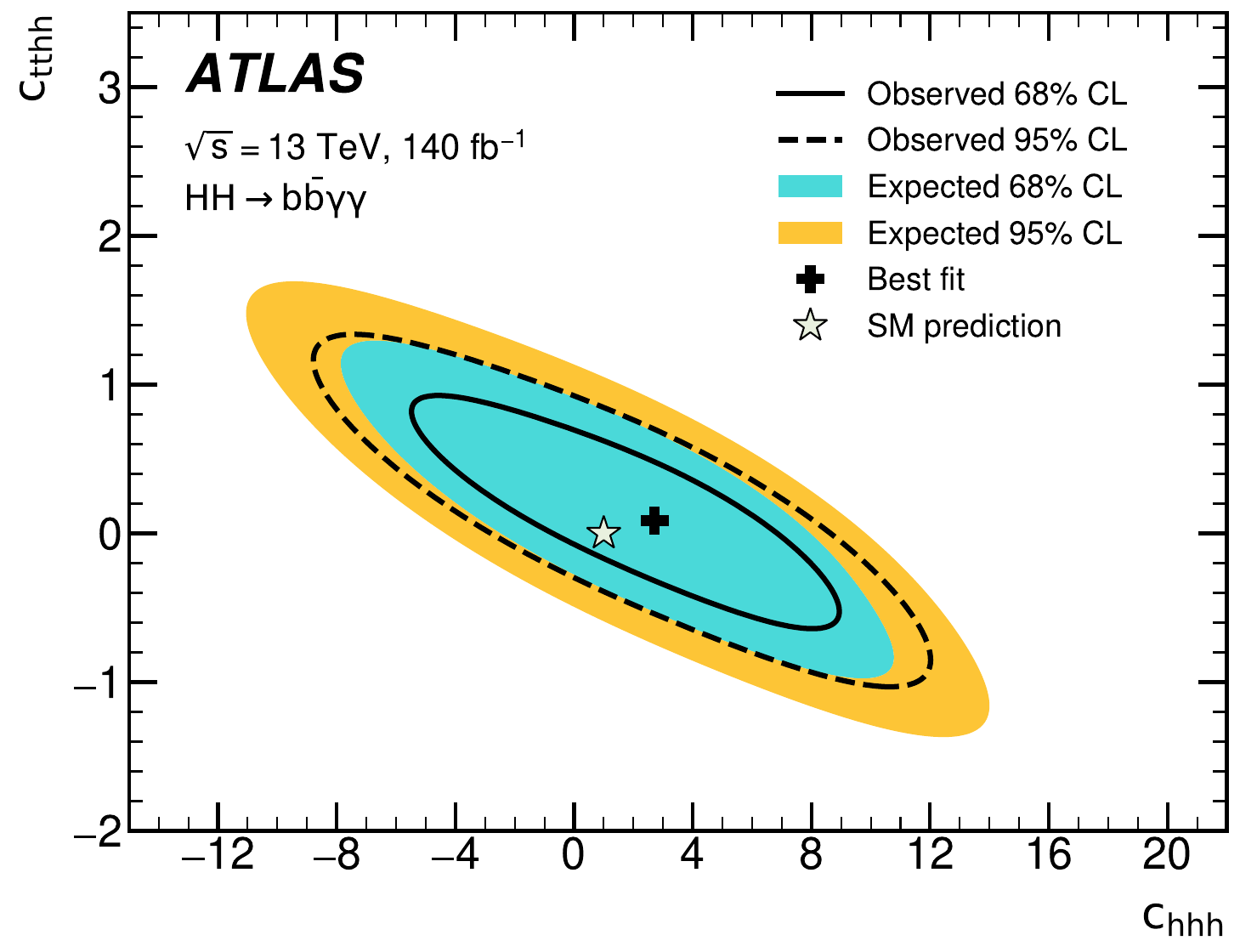}}\\
\caption{
Likelihood contours at 68\% (solid line) and 95\% (dashed line)
CL in the (a) $c_{gghh}$ versus $c_{hhh}$ and (b) $c_{tthh}$ versus
$c_{hhh}$ HEFT parameter space, with the remaining coefficient
fixed to its SM value.
The corresponding expected contours are shown by the
inner and outer shaded regions.
The SM prediction is indicated by the star, while
the best-fit value is denoted by the cross.
}
\label{fig:HEFT:2Dexclusions_withLogL}
\end{figure}

In the HEFT interpretation, constraints on the
coefficients $c_{hhh}$, $c_{tthh}$, and $c_{gghh}$ that describe
Higgs boson self-interactions as well as effective $t\bar{t}HH$ and $ggHH$
interactions are determined from the data from one-dimensional
scans of the profile likelihood function as a function of the
coefficients.
The operators corresponding to these coefficients do not impact
single Higgs boson production and decay at tree level and their effect on
the resonant background and on the Higgs boson branching ratios
is therefore neglected.
The one-dimensional constraints on the three HEFT coefficients
$c_{hhh}$, $c_{tthh}$ and $c_{gghh}$ are summarised in
Table~\ref{tab:HEFT_1D_coeff_limits}.
The difference between the $c_{hhh}$ constraint and the $\kl$
constraint previously presented in
Figure~\ref{fig:kl_and_k2v_1d_likelihood_scans}
is mainly due to the lack of VBF production in the former.
In addition, the observed constraints are comparable with those of
Ref.~\cite{ATLAS4b}, when evaluated using the same statistical procedure
of Ref.~\cite{ATLAS4b}. The width of the allowed 95\% CL interval
for $c_{gghh}$ is 20\% narrower, while that of the $c_{tthh}$ interval
is the same.
Figure~\ref{fig:HEFT:2Dexclusions_withLogL} shows two-dimensional
profile log-likelihood contours for the simultaneous variation of the
($c_{gghh}, c_{hhh}$) and ($c_{tthh}, c_{hhh}$) HEFT coefficients,
with the remaining coefficient fixed to its SM value.
 
\begin{table*}[htb]
\centering
\caption{
The observed and expected 95\% CL constraints on the HEFT Wilson
coefficients, obtained from one-dimensional scans of the profile
log-likelihood assuming that all other Wilson
coefficients are fixed to their SM values. The contribution from VBF
$HH$ production is subdominant to that from ggF and is neglected.
}
\label{tab:HEFT_1D_coeff_limits}
\begin{tabular}{lcc}
\toprule
Wilson coefficient & 95\% CL Observed & 95\% CL Expected \\
\midrule
$c_{hhh}$  & $[-1.7\phantom{0}, 7.7\phantom{0}]$ & $[-3.4\phantom{0}, 8.9\phantom{0}]$\\
$c_{tthh}$ & $[-0.28, 0.73]$ & $[-0.48, 0.94]$\\
$c_{gghh}$ & $[-0.42, 0.52]$ & $[-0.59, 0.69]$\\
\bottomrule
\end{tabular}
\end{table*}
 
\noindent
In addition, upper limits are set on the Higgs boson pair production
cross-section for seven benchmark points~\cite{Alasfar:2023xpc}
corresponding to different values of the five coefficients $c_{hhh}$,
$c_{tthh}$, $c_{gghh}$, $c_{ggh}$, and $c_{tth}$,
where the latter two correspond to an effective Higgs--gluon interaction
and to the Higgs--top Yukawa interaction.
The impact of these coefficients on single Higgs boson production
and decay is expected to be small compared to the signal and is
thus neglected.
Defined in Table~\ref{tab:HEFT_benchmark_def}, the benchmark points describe representative signal kinematics and $m_{HH}$ shape
features, and have sensitivities that can vary significantly between
one point and another.
For example, benchmark 1 results in a very soft $m_{HH}$ distribution
while benchmark 5 produces a more SM-like $m_{HH}$ distribution with
an enhanced tail.
 
\begin{figure}[hbtp]
\centering
\includegraphics[width=0.5\textwidth]{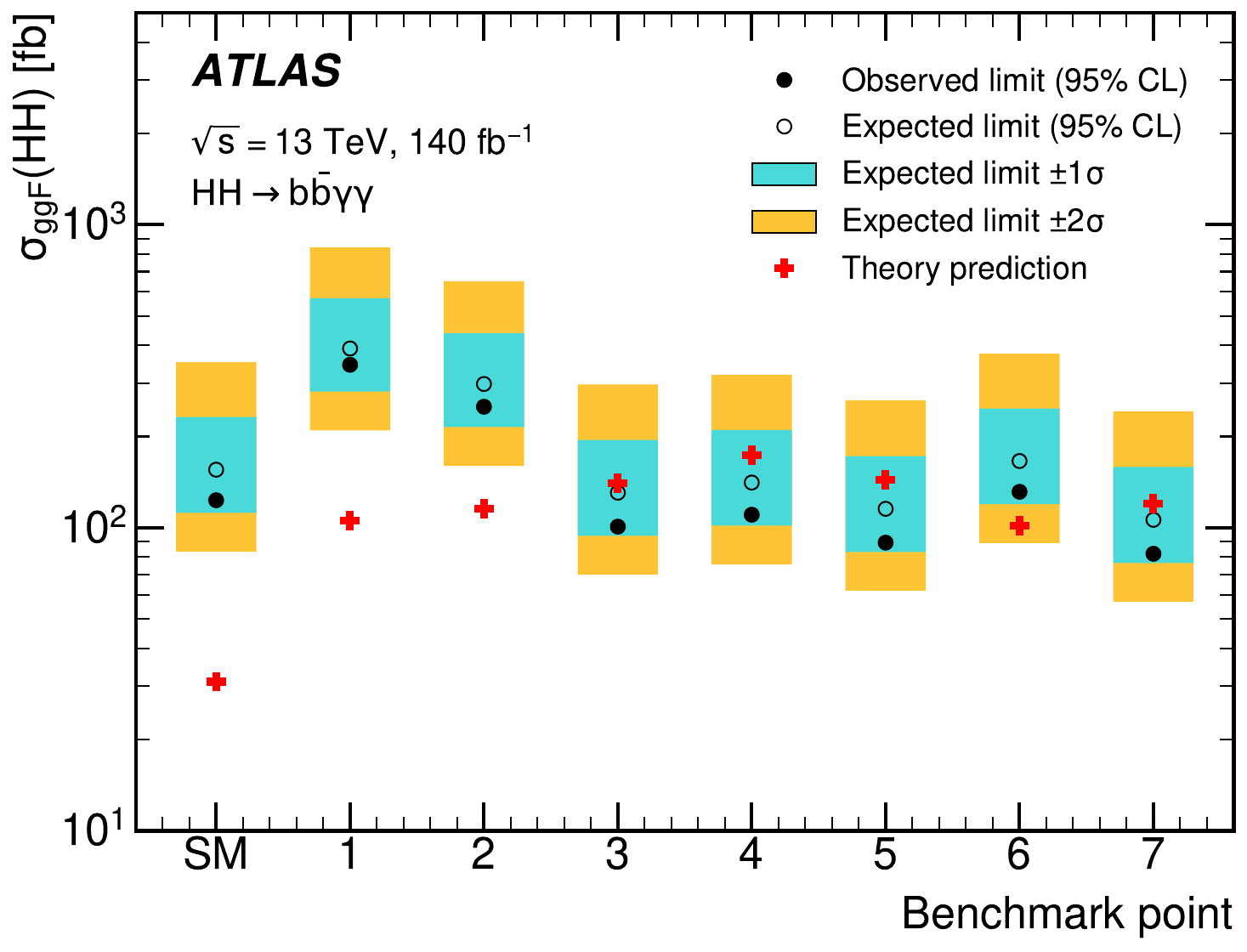}
\caption{The observed (filled circles) and expected (hollow
circles)
95\% CL upper limits on the $HH$ ggF production
cross-section in the SM and for seven HEFT
benchmark points defined in Ref.~\cite{Alasfar:2023xpc}.
The expected constraints are obtained from a background
hypothesis with $\sigma_{HH} =0$.
The predicted cross-sections of each of the models under consideration
are shown by the crosses.
Benchmarks where the filled circles are below the crosses are excluded.
The inner and outer shaded bands indicate the $\pm 1\sigma$
and $\pm 2\sigma$ variations on the expected limit
due to statistical and systematic uncertainties.
The contribution from VBF production to the total $HH$ production
cross-section is neglected.
}
\label{fig:HEFT_BM_xsec_UL}
\end{figure}

The resulting upper limits on the Higgs boson pair production cross-section
through gluon--gluon fusion are shown in
Figure~\ref{fig:HEFT_BM_xsec_UL}.
For benchmark points 3, 5 and 7, this analysis sets upper limits
similar to those set by the search for $HH\to 4b$
events~\cite{ATLAS4b}, and, in an analogous way, excludes these scenarios
at 95\% CL\@.
The remaining benchmarks (1, 2, 4, and 6) have updated definitions
compared to those used in Ref.~\cite{ATLAS4b} and therefore the results
cannot be directly compared.
Benchmark 4 is excluded for the first time at 95\% CL by this study,
while the other three scenarios are compatible with the data.
 
\begin{table*}[htb]
\centering
\caption{
The definitions of the seven HEFT benchmark points described in
Ref.~\cite{Alasfar:2023xpc}.
}
\label{tab:HEFT_benchmark_def} 
\begin{tabular}{lccccc}
\toprule
Benchmark & $c_{hhh}$ & $c_{tth}$ & $c_{ggh}$ & $c_{gghh}$ & $c_{tthh}$ \\
\midrule
SM &   $\phantom{-}1.00$            & 1.00  & $\phantom{-}0\phantom{/1}$ & $\phantom{-}0\phantom{/1}$ & $\phantom{-}0\phantom{/1}$ \\
1  &   $\phantom{-}5.11$            & 1.10  & $\phantom{-}0\phantom{/1}$ & $\phantom{-}0\phantom{/1}$ & $\phantom{-}0\phantom{/1}$ \\
2  &   $\phantom{-}6.84$            & 1.03  & $         - 1/3$           & $\phantom{-}0\phantom{/1}$ & $\phantom{-}1/6$ \\
3  &   $\phantom{-}2.21$            & 1.05  & $\phantom{-}1/2$           & $\phantom{-}1/2$           & $         - 1/3$ \\
4  &   $\phantom{-}2.79$            & 0.90  & $         - 1/3$           & $         - 1/2$           & $         - 1/6$ \\
5  &   $\phantom{-}3.95$            & 1.17  & $\phantom{-}1/6$           & $         - 1/2$           & $         - 1/3$ \\
6  & $-0.68$                        & 0.90  & $\phantom{-}1/2$           & $\phantom{-}1/4$           & $         - 1/6$ \\
7  & $-0.10$                        & 0.94  & $\phantom{-}1/6$           & $         - 1/6$           & $\phantom{-}1\phantom{/1}$ \\
\bottomrule
\end{tabular}
\end{table*}
 
\begin{figure}[htbp]
\centering
\includegraphics[width=0.45\textwidth]{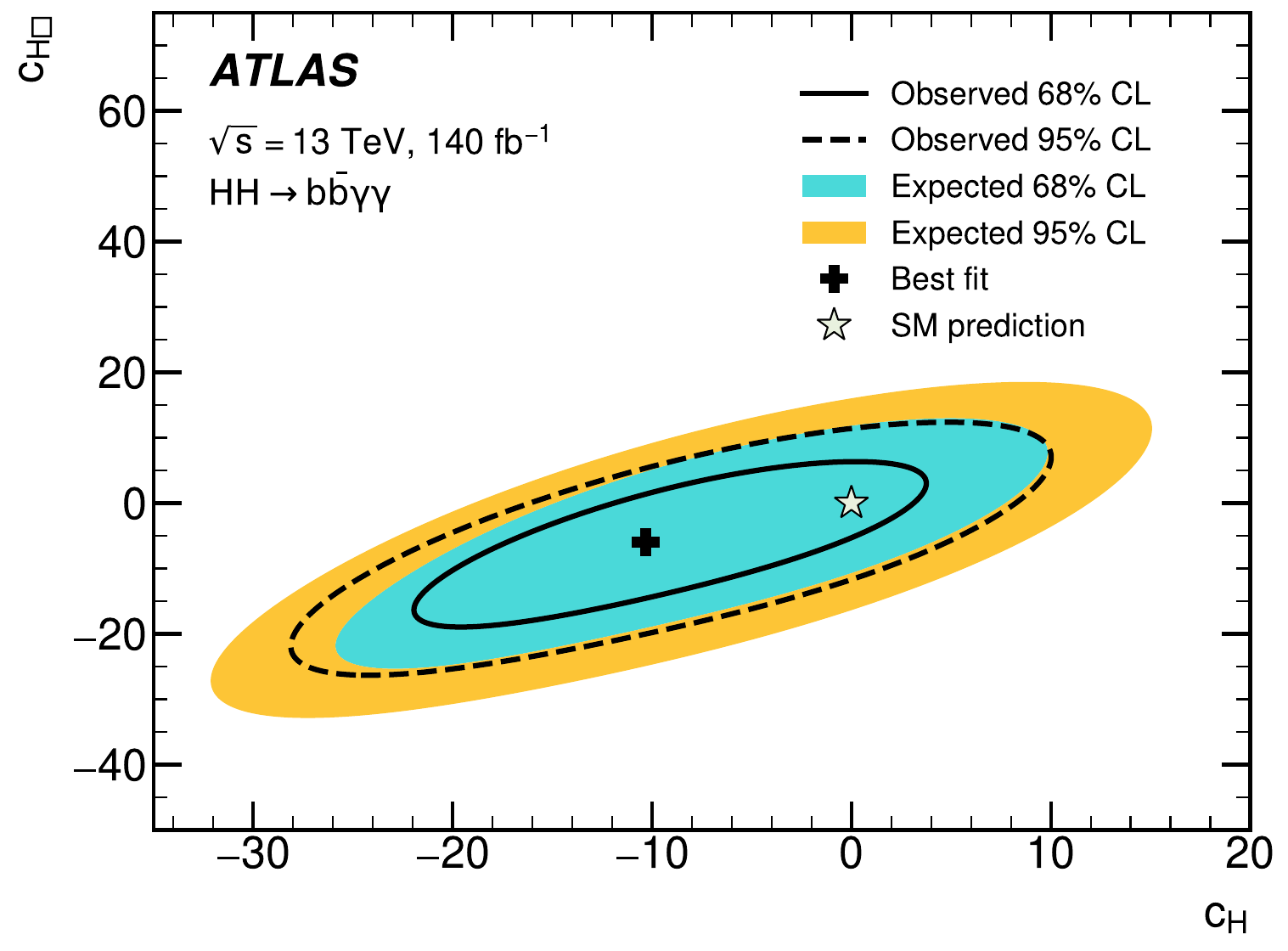}
\caption{
Likelihood contours at 68\% (solid line) and 95\% (dashed line)
CL in the $c_{{H}\boxed{}}$ versus $c_{H}$ SMEFT parameter space.
The corresponding expected contours are shown by the
inner and outer shaded regions.
The SM prediction is indicated by the star, while
the best-fit value is denoted by the cross.
}
\label{fig:eft_smeft_2d_likelihood_scans_linear_plus_quadratic}
\end{figure}

In the SMEFT interpretation, one-dimensional constraints are derived
on the Wilson coefficients after fixing all other coefficients to zero.
The results are obtained by including the contributions to the $HH$ and
$H$ cross-sections from both linear and quadratic terms in the Wilson
coefficient expansion.
An interpretation in which the expansion is truncated at linear order
is poorly constrained due to the dominance of the quadratic term and
can yield negative signal cross-sections. The impact of the operators
under study on ggF $HH$ production parameterised
as a function of $m_{HH}$ and on single Higgs boson production
parameterised as a function of the Higgs boson transverse momentum
are included in the interpretation. As in the case of HEFT, the
coefficients do not impact the Higgs boson decay branching ratios.
The one-dimensional constraints on the SMEFT Wilson coefficients in the scenario where
the other parameters are fixed to zero, as expected in the SM, are
summarised in Table~\ref{tab:SMEFT_1D_coeff_limits}.
When using the same statistical procedure of Ref.~\cite{ATLAS4b} to
determine the constraints on the SMEFT Wilson coefficients, the
results are only mildly affected, and the size of the 95\% CL interval
for $c_H$ ($c_{H\boxed{}}$) is 38\% (10\%) smaller
than that in Ref.~\cite{ATLAS4b}. Furthermore,
Figure~\ref{fig:eft_smeft_2d_likelihood_scans_linear_plus_quadratic}
shows two-dimensional likelihood scans as a function of the couplings
$c_{{H}\boxed{}}$ and $c_{H}$.

\begin{table*}[htb]
\centering
\caption{
The observed and expected 95\% CL constraints on the SMEFT Wilson
coefficients, obtained from one dimensional scans of the profile
log-likelihood assuming that all other Wilson
coefficients are fixed to their SM values.
The contribution from VBF production is neglected.
}
\label{tab:SMEFT_1D_coeff_limits}
\begin{tabular}{lcc}
\toprule
Wilson coefficient & 95\% CL Observed & 95\% CL Expected \\
\midrule
$c_{H}$          & $[-14.4, \phantom{0}6.2]$ & $[-16.8, \phantom{0}9.7]$ \\
$c_{{H}\boxed{}}$ & $[-\phantom{0}9.4, 10.2]$ & $[-12.4, 13.7]$ \\
\bottomrule
\end{tabular}
\end{table*}

\FloatBarrier

\section{Conclusion}
\label{sec:conclusion}
 
An updated search for nonresonant Higgs boson pair production
in the \yybb{} final state is performed using
the full Run~2 ATLAS data, corresponding to 140~fb$^{-1}$
of 13~\TeV{} $pp$ collisions.
The results supersede and expand upon those of a previous nonresonant search
based on the same data sample.
Compared to the previous publication, the classification
of events in orthogonal event categories is reoptimised
to increase the sensitivity to $HH$ production in the main
production modes, ggF and VBF, and to
the Higgs boson self-coupling and quartic coupling to $W,Z$ bosons.
The sensitivity is increased by 6\%--17\% depending on the parameter of interest.
The statistical procedure
for the interpretation of the observed yields in terms of the
signal coupling strength modifiers has also been updated.
In addition, the results are interpreted in the context
of the Higgs and SM effective field theory frameworks to constrain the
Wilson coefficients of operators describing anomalous Higgs
boson interactions.
 
No evidence of signal is found. In the most sensitive categories
of the analysis a small deficit of events in the signal
region leads to a 95\% CL upper limit on the $HH$ production
signal strength $\mu_{HH}<4.0$ that is lower than the
expected value of 5.0 (6.4) in the background-only $\mu_{HH}=0$
(SM $\mu_{HH}=1$) hypothesis.
The corresponding observed (expected) one-dimensional intervals
at 95\% CL for the self-coupling modifier $\kl$ and the quartic
coupling modifier \kvv are
$-1.4 < \kl < 6.9$ ($-2.8 < \kl < 7.8$)
and
$-0.5 < \kvv < 2.7$ ($-1.1 < \kvv < 3.3$), respectively.
From these results, one-dimensional limits on the Wilson coefficients of operators affecting Higgs
boson pair production in the Higgs effective field theory
($c_{hhh}, c_{tthh}, c_{gghh}$) and SM effective field theory
($c_{H}, c_{{H}\boxed{}}$) frameworks are inferred. In the former, the comparison
between the predicted gluon--gluon fusion $HH$ cross-sections and
the corresponding upper limits set by the analysis
excludes four of the seven benchmark points considered
at 95\% CL\@. While three of these were already excluded by
a similar interpretation of the results in the ATLAS search for $HH$ production
in the $4b$ final state, one newly proposed benchmark
is excluded for the first time by the results presented in this paper.
The one-dimensional constraints on the Wilson coefficients considered
in this analysis 
are up to 38\% tighter than those reported previously by ATLAS when evaluated using the same statistical procedure.


\section*{Acknowledgements}


We thank CERN for the very successful operation of the LHC, as well as the
support staff from our institutions without whom ATLAS could not be
operated efficiently.
 
We acknowledge the support of
ANPCyT, Argentina;
YerPhI, Armenia;
ARC, Australia;
BMWFW and FWF, Austria;
ANAS, Azerbaijan;
CNPq and FAPESP, Brazil;
NSERC, NRC and CFI, Canada;
CERN;
ANID, Chile;
CAS, MOST and NSFC, China;
Minciencias, Colombia;
MEYS CR, Czech Republic;
DNRF and DNSRC, Denmark;
IN2P3-CNRS and CEA-DRF/IRFU, France;
SRNSFG, Georgia;
BMBF, HGF and MPG, Germany;
GSRI, Greece;
RGC and Hong Kong SAR, China;
ISF and Benoziyo Center, Israel;
INFN, Italy;
MEXT and JSPS, Japan;
CNRST, Morocco;
NWO, Netherlands;
RCN, Norway;
MEiN, Poland;
FCT, Portugal;
MNE/IFA, Romania;
MESTD, Serbia;
MSSR, Slovakia;
ARRS and MIZ\v{S}, Slovenia;
DSI/NRF, South Africa;
MICINN, Spain;
SRC and Wallenberg Foundation, Sweden;
SERI, SNSF and Cantons of Bern and Geneva, Switzerland;
MOST, Taiwan;
TENMAK, T\"urkiye;
STFC, United Kingdom;
DOE and NSF, United States of America.
In addition, individual groups and members have received support from
BCKDF, CANARIE, Compute Canada and CRC, Canada;
PRIMUS 21/SCI/017 and UNCE SCI/013, Czech Republic;
COST, ERC, ERDF, Horizon 2020, ICSC-NextGenerationEU and Marie Sk{\l}odowska-Curie Actions, European Union;
Investissements d'Avenir Labex, Investissements d'Avenir Idex and ANR, France;
DFG and AvH Foundation, Germany;
Herakleitos, Thales and Aristeia programmes co-financed by EU-ESF and the Greek NSRF, Greece;
BSF-NSF and MINERVA, Israel;
Norwegian Financial Mechanism 2014-2021, Norway;
NCN and NAWA, Poland;
La Caixa Banking Foundation, CERCA Programme Generalitat de Catalunya and PROMETEO and GenT Programmes Generalitat Valenciana, Spain;
G\"{o}ran Gustafssons Stiftelse, Sweden;
The Royal Society and Leverhulme Trust, United Kingdom.
 
The crucial computing support from all WLCG partners is acknowledged gratefully, in particular from CERN, the ATLAS Tier-1 facilities at TRIUMF (Canada), NDGF (Denmark, Norway, Sweden), CC-IN2P3 (France), KIT/GridKA (Germany), INFN-CNAF (Italy), NL-T1 (Netherlands), PIC (Spain), ASGC (Taiwan), RAL (UK) and BNL (USA), the Tier-2 facilities worldwide and large non-WLCG resource providers. Major contributors of computing resources are listed in Ref.~\cite{ATL-SOFT-PUB-2023-001}.


\printbibliography
 
\clearpage
 
\begin{flushleft}
\hypersetup{urlcolor=black}
{\Large The ATLAS Collaboration}

\bigskip

\AtlasOrcid[0000-0002-6665-4934]{G.~Aad}$^\textrm{\scriptsize 102}$,
\AtlasOrcid[0000-0002-5888-2734]{B.~Abbott}$^\textrm{\scriptsize 120}$,
\AtlasOrcid[0000-0002-1002-1652]{K.~Abeling}$^\textrm{\scriptsize 55}$,
\AtlasOrcid[0000-0001-5763-2760]{N.J.~Abicht}$^\textrm{\scriptsize 49}$,
\AtlasOrcid[0000-0002-8496-9294]{S.H.~Abidi}$^\textrm{\scriptsize 29}$,
\AtlasOrcid[0000-0002-9987-2292]{A.~Aboulhorma}$^\textrm{\scriptsize 35e}$,
\AtlasOrcid[0000-0001-5329-6640]{H.~Abramowicz}$^\textrm{\scriptsize 151}$,
\AtlasOrcid[0000-0002-1599-2896]{H.~Abreu}$^\textrm{\scriptsize 150}$,
\AtlasOrcid[0000-0003-0403-3697]{Y.~Abulaiti}$^\textrm{\scriptsize 117}$,
\AtlasOrcid[0000-0002-8588-9157]{B.S.~Acharya}$^\textrm{\scriptsize 69a,69b,m}$
% ...
\bigskip
\\

% ...
$^{29}$Physics Department, Brookhaven National Laboratory, Upton NY; United States of America.\\
% ...
$^{35}$$^{(a)}$Facult\'e des Sciences Ain Chock, R\'eseau Universitaire de Physique des Hautes Energies - Universit\'e Hassan II, Casablanca;$^{(b)}$Facult\'{e} des Sciences, Universit\'{e} Ibn-Tofail, K\'{e}nitra;$^{(c)}$Facult\'e des Sciences Semlalia, Universit\'e Cadi Ayyad, LPHEA-Marrakech;$^{(d)}$LPMR, Facult\'e des Sciences, Universit\'e Mohamed Premier, Oujda;$^{(e)}$Facult\'e des sciences, Universit\'e Mohammed V, Rabat;$^{(f)}$Institute of Applied Physics, Mohammed VI Polytechnic University, Ben Guerir; Morocco.\\
% ...
$^{49}$Fakult\"{a}t Physik , Technische Universit{\"a}t Dortmund, Dortmund; Germany.\\
% ...
$^{55}$II. Physikalisches Institut, Georg-August-Universit\"{a}t G\"ottingen, G\"ottingen; Germany.\\
% ...
$^{69}$$^{(a)}$INFN Gruppo Collegato di Udine, Sezione di Trieste, Udine;$^{(b)}$ICTP, Trieste;$^{(c)}$Dipartimento Politecnico di Ingegneria e Architettura, Universit\`a di Udine, Udine; Italy.\\
% ...
$^{102}$CPPM, Aix-Marseille Universit\'e, CNRS/IN2P3, Marseille; France.\\
% ...
$^{117}$Department of Physics, New York University, New York NY; United States of America.\\
% ...
$^{120}$Homer L. Dodge Department of Physics and Astronomy, University of Oklahoma, Norman OK; United States of America.\\
% ...
$^{150}$Department of Physics, Technion, Israel Institute of Technology, Haifa; Israel.\\
$^{151}$Raymond and Beverly Sackler School of Physics and Astronomy, Tel Aviv University, Tel Aviv; Israel.\\
% ...
$^{a}$ Also Affiliated with an institute covered by a cooperation agreement with CERN.\\
$^{b}$ Also at An-Najah National University, Nablus; Palestine.\\
$^{c}$ Also at Borough of Manhattan Community College, City University of New York, New York NY; United States of America.\\
$^{d}$ Also at Center for High Energy Physics, Peking University; China.\\
$^{e}$ Also at Center for Interdisciplinary Research and Innovation (CIRI-AUTH), Thessaloniki; Greece.\\
$^{f}$ Also at Centro Studi e Ricerche Enrico Fermi; Italy.\\
% ...
$^{m}$ Also at Department of Physics, King's College London, London; United Kingdom.\\
% ...

\end{flushleft}


\end{document}